\documentclass[review,sort&compress]{elsarticle}
\usepackage{amsmath}
\usepackage{amsfonts}
\usepackage[nolists,tablesfirst]{endfloat}
\usepackage{epstopdf}
\usepackage{multicol}
\usepackage{multirow}
\usepackage{setspace}
\usepackage{rotating}
\usepackage{dcolumn}
\usepackage{caption}
\newcolumntype{d}{D{.}{.}{2.2}}
\usepackage{supertabular}
\usepackage[colorlinks=true,citecolor=blue,linkcolor=blue]{hyperref}%

\begin{document}
\begin{frontmatter}
\journal{Acta Materialia}
\title{Thermodynamic Stability of Mg-based Ternary Long-Period Stacking Ordered Structures}
\author{James E. Saal\corref{cor1}}
\ead{j-saal@northwestern.edu}
\author{C. Wolverton\corref{dummy}}

\address{Department of Materials Science and Engineering,\\
Northwestern University, Evanston, IL 60201, USA}
\cortext[cor1]{Corresponding author}

\begin{abstract}
Mg alloys containing long-period stacking ordered (LPSO) structures exhibit remarkably high tensile yield strength and ductility. They have been found in a variety of ternary Mg systems of the general form Mg-X$\rm^L$-X$\rm^S$, where X$\rm^L$ and X$\rm^S$ are elements larger and smaller than Mg, respectively. In this work, we examine the thermodynamic stability of these LPSO precipitates with density functional theory, using a newly proposed structure model based on the inclusion of a Mg interstitial atom. We predict the stabilities for 14H and 18R LPSO structures for many Mg-X$\rm^L$-X$\rm^L$ ternary systems: 85 systems consisting of X$\rm^L$=rare earths (RE) Sc,Y,La-Lu and X$\rm^S$=Zn,Al,Cu,Co,Ni.  We predict thermodynamically stable LPSO phases in all systems where LPSO structures are observed. In addition, we predict several stable LPSO structures in new, as-yet-unobserved Mg-RE-X$\rm^S$ systems. Many non-RE X$\rm^L$ elements are also explored on the basis of size mismatch between Mg and X$\rm^L$, including Tl,Sb,Pb,Na,Te,Bi,Pa,Ca,Th,K,Sr --- an additional 55 ternary systems. X$\rm^L$=Ca, Sr, and Th are predicted to be most promising to form stable LPSO phases, particularly with X$\rm^S$=Zn. Lastly, several previously observed trends amongst known X$\rm^L$ elements are examined. We find that favorable mixing energy between Mg and X$\rm^L$ on the FCC lattice and the size mismatch together serve as excellent criteria determining X$\rm^L$ LPSO formation.
\end{abstract}
\end{frontmatter}

\section{Introduction}
Mg-based alloys are often considered potential lightweight structural alloys for transportation applications in efforts to improve efficiency. However, poor mechanical strength and ductility have long been impediments to wide industrial use of Mg. In 2001, Kawamura et al. reported a Mg alloy with a nominal composition of Mg$_{97}$Zn$_1$Y$_1$ exhibiting a remarkably high tensile yield strength of 610 MPa at 16\% elongation\cite{Kawamura2001}. This strength is coupled with the appearance of a novel ternary precipitate exhibiting order with long periods along the c-axis of the HCP Mg matrix\cite{Inoue2011}. Referred to as long period stacking ordered (LPSO) structures, these precipitates, and their resulting high strength, have since been observed in a variety of ternary Mg systems\cite{Amiya2003,Yamasaki2005,Kawamura2006,YAMADA2006,Kawamura2007,Itoi2008,Nie2008,Yokobayashi2011,Mi2013,Jin2013,Leng2013}. However, LPSO systems typically contain at least 1 at.\% rare earth (RE) elements, making such alloys prohibitively expensive for high-volume industrial applications.

LPSO structures have been reported in several Mg-X$\rm^L$-X$\rm^S$ ternary systems, consisting of Mg, an alloying element larger than Mg (X$\rm^L$), and an alloying element smaller than Mg (X$\rm^S$). Currently, the following alloying elements have been reported to form LPSO structures: X$\rm^L$=Y,Gd,Tb,Dy,Ho,Er,Tm and X$\rm^S$=Zn,Al,Cu,Co,Ni\cite{Amiya2003,Yamasaki2005,Kawamura2006,YAMADA2006,Kawamura2007,Itoi2008,Nie2008,Yokobayashi2011,Mi2013,Jin2013,Leng2013}. Of the 35 possible Mg-X$\rm^L$-X$\rm^S$ ternaries listed above, only 11 have reported LPSO formation, as summarized in Figure \ref{fig:stabmap}. This is likely a consequence of a focus on $\rm^L$=Zn systems, particularly the Mg-Y-Zn system.

An important issue in predicting the properties for these LPSO structures with atomistic models such as density functional theory (DFT) is that their precise structure has remained elusive. The LPSO structure is readily observed to contain long range order along the c-axis. For many systems, the structure is, at first, of rhombohedral symmetry with 18 atomic layers per repeat unit (the 18R structure) and, after annealing, transforms into a different structure with hexagonal symmetry and 14 layers per repeat unit (the 14H structure)\cite{Itoi2004,Kawamura2007}.  However, it was only recently that complete structure models were proposed. The first, reported by Zhu et al. for Mg-Y-Zn\cite{Zhu2010}, consists of alternating groups of HCP-stacked Mg planes (five planes for 14H and four for 18R) and pairs of ternary FCC-stacked Mg-Y-Zn planes. Importantly, the ternary planes were reported to be fully ordered chemically. The second structural model, reported by Egusa and Abe, for Mg-Y-Zn and Mg-Er-Zn\cite{Egusa2012}, modified the 2011 model, replacing the HCP Mg planes adjacent to the FCC ternary planes with binary Mg-X$\rm^L$ planes, with the change in composition along the c-axis more gradual.  In both models, the only difference between the 14H and 18R structures is the quantity of HCP Mg planes.

In our previous work\cite{Saal2012}, we compared the DFT-predicted thermodynamic stability of the two structure models\cite{Zhu2010,Egusa2012} in the Mg-Y-Zn system and found that (1) for both 14H and 18R structures the newer, ``gradual'' model of Egusa and Abe\cite{Egusa2012} was energetically preferred over the sharper one by Zhu et al.\cite{Zhu2010}, (2) the 14H structure is more stable than 18R, and (3) all of the calculated structures were metastable (i.e. they were higher in energy than combinations of other phases in the Mg-Y-Zn system). Hence, using these structural models, DFT calculations predict that LPSO structures are not thermodynamic ground states. This conclusion could be reconciled with the experimental observations of LPSO structures in one of two ways: either (i) the observed LPSO structures are, in fact, metastable, or (ii) these previous structural models or DFT calculations are incorrect.

Since our previous work, a third structure model has been proposed\cite{Egusa-PC}. It has been observed that the ``gradual'' LPSO structure model, when relaxed in DFT, expands in such a way as to create a large interstitial site within the ternary ordered layers\cite{Egusa-PC}. DFT calculations for this ``interstitial'' LPSO model in the Mg-Y-Zn system have shown that this structure is thermodynamically stable relative to the other phases\cite{Egusa-PC}. Another recent work\cite{Ma2012} examined with DFT various Y-Zn ordered clusters in the LPSO ternary layers, including interstitials. The stability of the interstitial model suggests this structure is more accurate than the previous sharp and gradual structure models and we can use this model to test the stability of a wide variety of LPSO precipitate chemistries in DFT.

In an effort to discover more affordable non-RE alloying elements which can form LPSO structures, we employ DFT calculations to predict the stability of LPSO structures in every LPSO-forming ternary system to examine the effect of chemistry on LPSO stability. We then extend the study to explore the possibility of novel non-RE elements capable of forming LPSO structures. We begin by exploring the thermodynamic stability of the interstitial LPSO structure model with DFT in detail for the Mg-Y-Zn system, finding that including Mg interstitials promotes the stability of the structure over the older models. We then systematically predict the stability of the interstitial LPSO structure in 85 RE-containing Mg-X$\rm^L$-X$\rm^S$ ternary systems, for X$\rm^L$=RE (Sc,Y,La-Lu) and X$\rm^S$=Zn,Al,Cu,Co,Ni. For the 11 systems where LPSO phases are observed, our calculations predict all of these phases to be thermodynamically stable.  In addition, we predict 41 stable RE-containing LPSO phases in systems where they have not been currently reported.  These 41 as-yet-unobserved LPSO phases represent predictions awaiting experimental confirmation. From these results, we test the validity of previously proposed rules for LPSO forming systems, including the effect of the size of the X$\rm^L$ element and the mixing energy between Mg and X$\rm^L$ on the FCC lattice. These design rules are then used to predict several candidate non-RE X$\rm^L$ elements that may also form LPSO structures, which we then calculate with DFT. We predict the stability of LPSO for 55 non-RE-containing systems and find several systems for which LPSO phases are low in energy, competitive with thermodynamic stability.  From these calculations, we predict that X$\rm^L$=Ca, Sr, and Th are promising LPSO forming elements.

\section{Methodology}\label{sec:method}
DFT calculations are performed with the Vienna Ab-initio Simulation Package (VASP)\cite{Kresse1996b,Kresse1996}, employing the projected augmented wave method potentials\cite{Kresse1999b} and the exchange and correlation functional of Perdew, Burke, and Ernzerhof\cite{Perdew1996}. All degrees of freedom for the crystal structures are relaxed, including volume, cell shape, and internal atomic coordinates, to determine the 0K energetic ground state structure. An energy cutoff of 520 eV and gamma-centered k-point meshes of around 8000 k-points per reciprocal atom are used in the relaxation. k-space integration is performed by the first-order Methfessel-Paxton approach with a smearing width of 0.2 eV during structural relaxation and then by the tetrahedron method with Bl\"{o}chl corrections during a final, static calculation for accurate total energy. The f-electrons of the lanthanide elements were treated as core electrons, an approximation that has shown to produce accurate thermodynamic properties for lanthanide-containing structures\cite{Gao2007,Mao2011,Saal2011b,Issa2013}. Calculations for systems containing Co and Ni were spin polarized with an initialized ferromagnetic structure.

For an LPSO structure to be thermodynamically stable, it must be stable with respect to every combination of unary, binary, and ternary phases in its respective ternary system. We define the thermodynamic stability of an LPSO structure, $\rm{\Delta E_{stab}}$(LPSO), by:
\begin{equation}\label{eqn:stability}
\rm{\Delta E_{stab}(LPSO)=E(LPSO)-\sum_i N_i \mu_i}
\end{equation}
\noindent where E(x) is the DFT predicted total energy of structure x, N$\rm_i$ is the amount of element i, and $\mu\rm_i$ is the chemical potential of element i. To determine the set of $\mu\rm_i$ chemical potentials, we employ the following two facts: first, for a system in equilibrium, the chemical potential of each element must be the same in every stable phase; second, the total energy of a structure is simply the composition weighted sum of the constituent chemical potentials,
\begin{equation}\label{eqn:mu}
\rm{E(x)=\sum_i N_i \mu_i}
\end{equation}
From these points, we construct a linear system of equations where Equation \ref{eqn:mu} is defined for each stable phase at the LPSO structure composition (excluding the LPSO structure itself) and solve for each $\mu\rm_i$. The formation energy, $\rm{\Delta E_{F}}$, is defined similarly to $\rm{\Delta E_{stab}}$ and Equation \ref{eqn:stability}, but the $\mu\rm_i$ chemical potentials are determined from the elemental structures instead of the equilibrium structures.

To calculate the set of stable phases (i.e the convex hull), we have employed the Open Quantum Materials Database (OQMD)\cite{Saal2013}, a high-throughput DFT database of total energies for every crystal structure found in the International Crystal Structure Database (ICSD)\cite{Bergerhoff1983,Belsky2002} with primitive cells less than 30 atoms and without partial site occupancy. For the 140 Mg-X$\rm^L$-X$\rm^S$ ternary systems examined in this work, this amounts to DFT calculations of over 3900 compounds. From this database of compounds, the most stable set of structures at a given composition, from which $\mu\rm_i$ are determined in Equation \ref{eqn:stability}, are calculated by grand canonical linear programming (GCLP)\cite{Wolverton2007a,Saal2013,Akbarzadeh2007,Kirklin2013}. With GCLP, since both the composition and the free energy are linear as a function of quantity of different phases in a system, the set of phases that has the minimum total free energy at a given composition can be determined by linear programming.

To illustrate the application of Equation \ref{eqn:stability}, the phases that are stable, excluding the LPSO structures, at the 14H-i Mg$_{71}$Y$_{8}$Zn$_{6}$ LPSO composition are Mg, MgYZn, and Mg$_3$Y (as listed in Table \ref{tab:Mg-RE-Zn}). The 14H-i structure will be defined in Section \ref{sec:xtal}. By Equation \ref{eqn:stability}, the stability of the 14H-i Mg$_{71}$Y$_{8}$Zn$_{6}$ LPSO structure is the energy of the LPSO relative to the composition-weighted sum of the competing phases:
\begin{equation}\label{eqn:stability}
\rm{\Delta E_{stab}(Mg_{71}Y_{8}Zn_{6})=E(Mg_{71}Y_{8}Zn_{6})-59E(Mg)-6E(MgYZn)-2E(Mg_3Y)}
\end{equation}
\noindent The energy of this reaction, also given in Table \ref{tab:Mg-RE-Zn}, is -12 meV/atom, where the negative value indicates the phase is stable. In other words, the 14H-i Mg$_{71}$Y$_{8}$Zn$_{6}$ LPSO structure is a stable phase as it lies 12 meV/atom below the convex hull composed of Mg, MgYZn, and Mg$_3$Y.

It should be noted that the predicted stabilities are subject to the availability of crystal structures in the ICSD. For example, some of the experimentally observed ternary phases in the Mg-Y-Zn system (W-Mg$_3$Y$_2$Zn$_3$, Z-Mg$_{28}$Y$_7$Zn$_{65}$, I-Mg$_3$YZn$_6$, H-Mg$_{15}$Y$_{15}$Zn$_{70}$, X-Mg$_{12}$YZn)\cite{Shao2006,Hamaya2013} do not have fully determined structures in the ICSD, so they are not included in our study. Therefore, the convex hull energetics in this work should be consider an upper bound on the true convex hull (i.e. the convex hull energies could be lower than those in the current work but not higher). Consequently, the DFT stabilities for the LPSO structures in this work are a lower bound (i.e. the stability could be more positive but not more negative than currently predicted).

Although computationally demanding and outside the scope of the current work, crystal structure prediction tools\cite{Oganov2006,Meredig2012,Trimarchi2007} can be applied to such systems with unknown compounds to at least approximate the formation energies of the phases in a given ternary competing with the LPSO structure. We approach the problem of unexplored systems and structures by calculating simple ordered structures in the FCC, BCC, and HCP lattices for all systems in this work. The included simple structures consist of binary compounds (L1$_2$, L1$_0$, D0$_3$, B2, B$\rm_h$, and D0$_{19}$) and the ternary X$_2$YZ Heusler compound. In this way, these prototype structures, although likely not on the true convex hull, may provide a better approximation for the convex hull energy in systems where experimentally determined crystal structures data may not be available. In other words, a predicted convex hull energy which includes a prototype will be more negative than without the prototype and closer to the true value. It appears this is an important consideration for the Mg-X$\rm^L$-X$\rm^S$ ternaries considered in this work since most of their convex hulls from the OQMD at LPSO compositions contain prototypes. The sets of stable phases at every LPSO composition are given in Tables \ref{tab:Mg-RE-Zn}-\ref{tab:Mg-RE-Ni}.

\section{Results and Discussion}
\subsection{Comparison of LPSO Structure Models}\label{sec:xtal}
\begin{figure}[tbp]
\centering %
\includegraphics[width=3in]{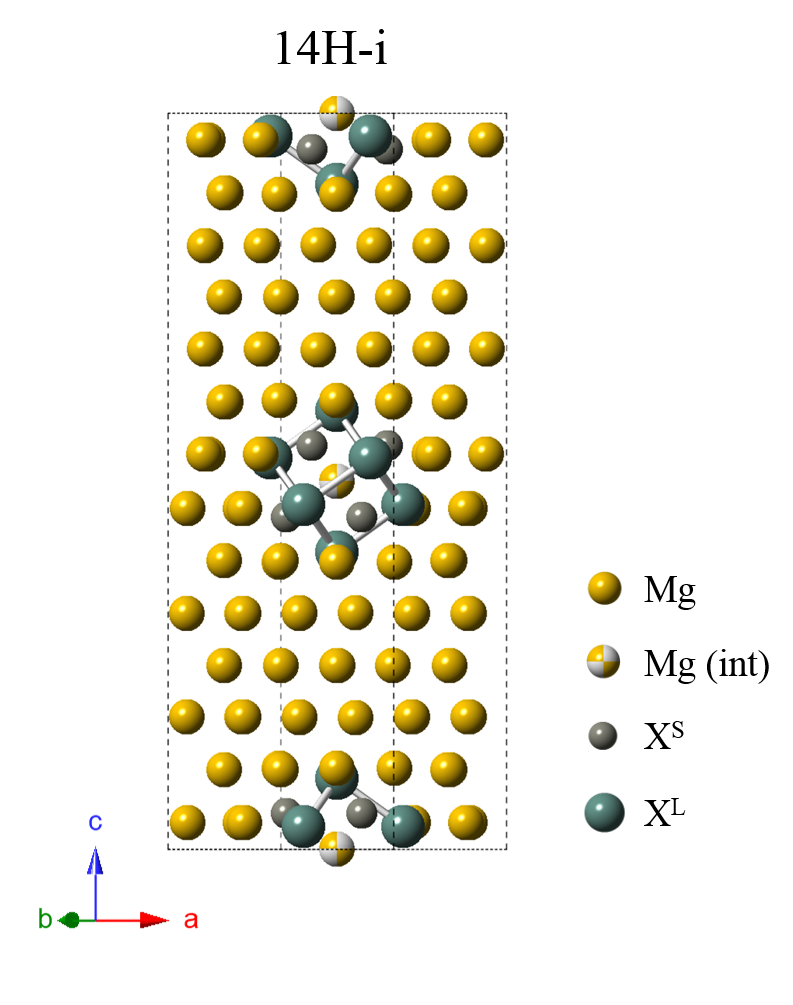}\\
\caption{The Mg$_{71}$X$\rm^L_{8}$X$\rm^S_{6}$ 14H-i LPSO crystal structure. A full X$\rm^S_{6}$X$\rm^L_{8}$ L1$_2$-arranged cluster can be seen in the middle of the cell with a Mg interstitial site at the center. Note that the origin has been shifted by 0.5,0.5,0 with respect to coordinates in Table \ref{tab:14Hi-structure}.}
\label{fig:crystal}
\end{figure}

The 14H and 18R gradual LPSO structures by Egusa and Abe\cite{Egusa2012} have stoichiometries of Mg$_{70}$X$\rm^L_{8}$X$\rm^S_{6}$ and Mg$_{58}$X$\rm^L_{8}$X$\rm^S_{6}$, respectively. The arrangement of the eight X$\rm^L$ and six X$\rm^S$ atoms within the four FCC stacked binary and ternary layers of the gradual LPSO structure model unit cell forms an X$\rm^S_{6}$X$\rm^L_{8}$ L1$_2$-arranged cluster in the Mg matrix, as shown in Figure \ref{fig:crystal} for 14H. Egusa and Abe\cite{Egusa2012} noted significant displacement of the X$\rm^L$ and X$\rm^S$ atoms in this cluster occurred after DFT relaxation of the ideal structure, with the X$\rm^L$ atoms moving towards the center of the cluster and the X$\rm^S$ atoms moving away, reducing the X$\rm^S$-X$\rm^S$ interatomic distance. Later DFT work from the same authors showed that this relaxation creates a large interstitial site at the body center of the L1$_2$ cluster, and the inclusion of an interstitial atom on this site thermodynamically stabilizes the structure\cite{Egusa-PC}. Analysis of the Mg-Y-Zn 14H and 18R gradual structures from our current calculations confirm this relaxation. The minimum nearest neighbor distances about the interstitial site ($int$) in the body center of the L1$_2$ cluster in the 14H structure are 3.16 and 3.40 $\rm{\AA}$ for the $int$-Zn and $int$-Y distances, respectively, large enough for an interstitial atom to be included. This interstitial site is also indicated in Figure \ref{fig:crystal}. For comparison, the distance of the next largest interstitial site to a nearest neighbor is 2.25 $\rm\AA$, indicating that there exists only one large interstitial site in the gradual LPSO structure.

To test which species of interstitial atom (Mg, X$\rm^L$, or X$\rm^S$) is the most stable, we calculate the energy to insert interstitial atom i, $\rm{\Delta E_{int}^i}$, for the three possible interstitial species in the 14H interstitial Mg-Y-Zn structure, Mg$_{70}$Y$_{8}$Zn$_{6}$($int$), where $int$ is the interstitial atom:
\begin{equation}\label{eqn:14H-14Hi}
\rm{\Delta E_{int}^{Mg} = Mg_{70}Y_{8}Zn_{6}(Mg) - Mg_{70}Y_{8}Zn_{6}  - \mu_{Mg} = -1.864 eV/int}
\end{equation}
\begin{equation}
\rm{\Delta E_{int}^Y = Mg_{70}Y_{8}Zn_{6}(Y) - Mg_{70}Y_{8}Zn_{6}  - \mu_{Y} = -1.474 eV/int}
\end{equation}
\begin{equation}
\rm{\Delta E_{int}^{Zn} = Mg_{70}Y_{8}Zn_{6}(Zn) - Mg_{70}Y_{8}Zn_{6}  - \mu_{Zn} = -1.032 eV/int}
\end{equation}
\noindent For all three defect formation energies, the $\mu\rm_i$ elemental chemical potentials are determined from the same set of stable compounds in the Mg-Y-Zn system at the LPSO composition: Mg, MgYZn, and Mg$_3$Y. Note that the experimentally observed stable Mg-rich Mg-Y binary compound is Mg$_{24}$Y$_5$, but our DFT calculations predict Mg$_3$Y D0$_3$ as more stable, in agreement with previous calculations\cite{Tao2011}. Mg$_{24}$Y$_5$ lies 3 meV/atom above the DFT convex hull, an energy difference that does not qualitatively affect the results in this work. All three interstitial defect formation energies are negative, indicating that they each stabilize the 14H gradual structure with their presence. Mg interstitials are predicted to be preferred as they have the most favorable formation energy and, thus, produce the most stable LPSO structure with respect to the other phases in the Mg-Y-Zn ternary system.

We calculate $\Delta \rm{E_{int}^{Mg}}$ for the X$\rm^L$=RE and X$\rm^S$=Al,Zn LPSO systems, shown in Figure \ref{eqn:14H-14Hi}. All the $\Delta \rm{E_{int}^{Mg}}$ values are negative, indicating that the interstitial Mg atom promotes the stability of the LPSO structure, by as much as -2.109 eV/defect for the Mg-Gd-Al system. We also predict $\Delta \rm{E_{int}^{Mg}}$ for the 18R LPSO structure for a selection of ternary systems by:
\begin{equation}\label{eqn:18R-18Ri}
\rm{\Delta E_{int}^{Mg} = Mg_{58}X^L_8X^S_6(Mg) - Mg_{58}X^L_8X^S_6  - \mu(Mg)}
\end{equation}		
\noindent The resulting the 18R $\Delta \rm{E_{int}^{Mg}}$ values are given in parentheses, in eV/defect:  Mg-Gd-Zn(-1.846), Mg-Y-Cu(-1.6375), Mg-Y-Co(-1.698), Mg-Y-Ni(-1.623), Mg-Gd-Al(-2.137).  As with the 14H structures, Mg interstitials stabilize the 18R structure. Indeed, \emph{for every case in this work, the LPSO structure with the interstitial Mg atoms are more stable than their gradual model equivalent.} Our DFT calculations lend strong support to the validity of the interstitial structure model proposed by Egusa and Abe\cite{Egusa-PC}. Based on these results, we will focus the remainder of the work on the LPSO gradual structures containing Mg interstitials, hereafter referred to as 14H-i and 18R-i. The DFT relaxed Mg-Y-Zn 14H-i and 18R-i crystal structures are given in Tables \ref{tab:14Hi-structure} and \ref{tab:18Ri-structure}. The relaxed Mg-RE-X$\rm^S$ 14H-i and 18R-i crystal structure parameters are provided in Tables \ref{tab:params-zn}-\ref{tab:params-ni}.

\begin{figure}[tbp]
\centering %
\includegraphics[width=5in]{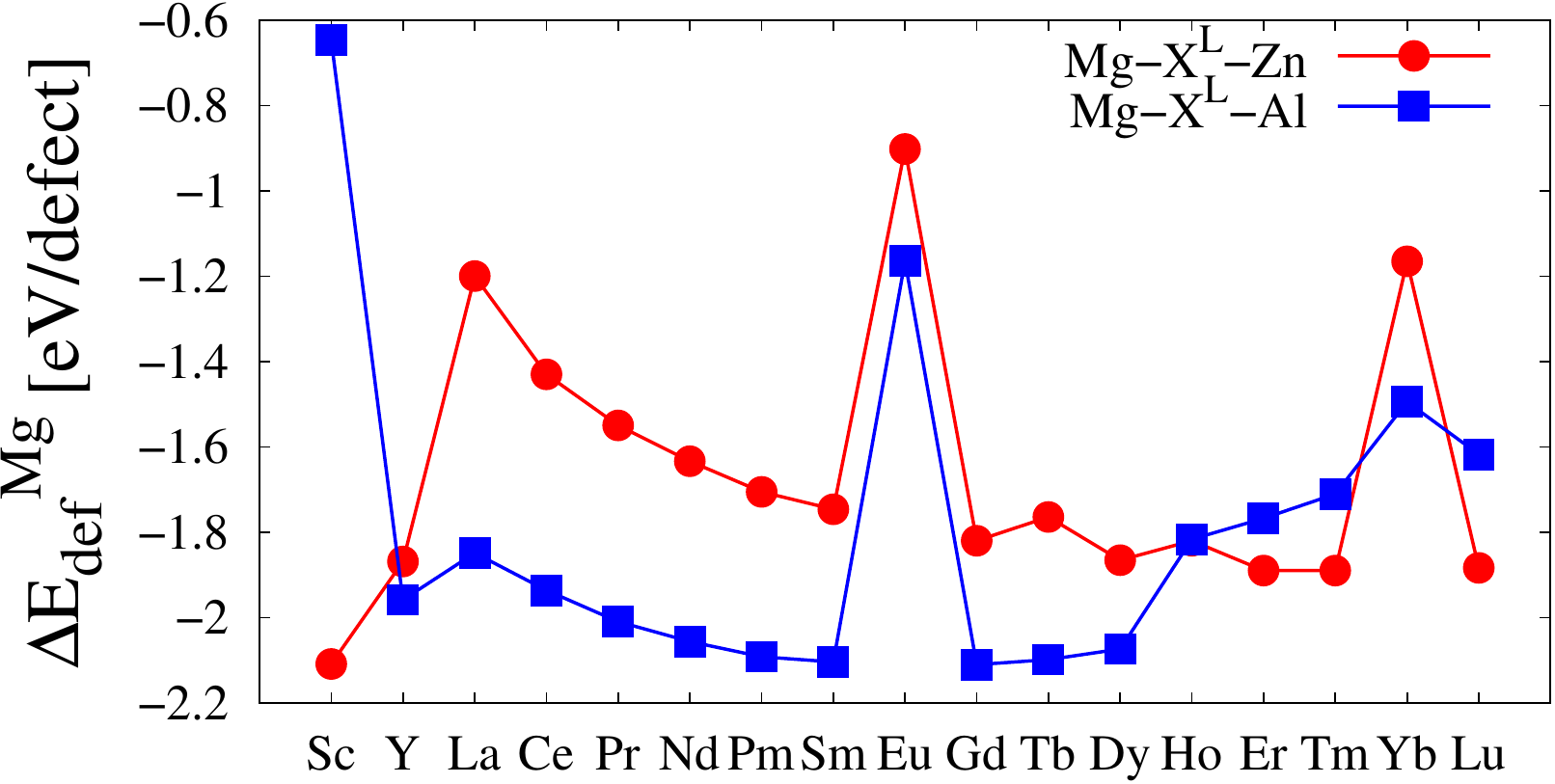}\\
\caption{DFT predicted Mg interstitial defect formation energy, $\Delta \rm{E_{int}^{Mg}}$, for the gradual 14H LPSO structures (Equation \ref{eqn:14H-14Hi}). Negative values indicate the interstitial Mg atom promotes the stability of the LPSO structures.}
\label{fig:14H-14Hi}
\end{figure}

\begin{table}[htbp]
  \centering
  \caption{DFT relaxed atomic positions for the Mg$_{71}$Y$_{8}$Zn$_{6}$ 14H-i LPSO structure, with spacegroup P6$_3$/mcm (193) and lattice parameters a=11.15$\rm\AA$ c=36.36$\rm\AA$.}
    \begin{tabular}{llccc}
        \hline\hline
Atom & site & x & y & z \\
\hline
    Mg1   & 24l   & 0.165 & 0.655 & 0.037 \\
    Mg2   & 24l   & 0.830 & 0.169 & 0.110 \\
    Mg3   & 24l   & 0.165 & 0.663 & 0.180 \\
    Mg4   & 12k   & 0.494 & 0.000 & 0.108 \\
    Mg5   & 12k   & 0.836 & 0.000 & 0.179 \\
    Mg6   & 12k   & 0.329 & 0.000 & 0.180 \\
    Mg7   & 12j   & 0.168 & 0.332 & 0.250 \\
    Mg8   & 8h    & 0.333 & 0.667 & 0.108 \\
    Mg9   & 6g    & 0.498 & 0.000 & 0.250 \\
    Mg10  & 4c    & 0.333 & 0.667 & 0.250 \\
    Mg11  & 2a    & 0.000 & 0.000 & 0.250 \\
 Mg12 int & 2b    & 0.000 & 0.000 & 0.000 \\
    Zn    & 12k   & 0.777 & 0.000 & 0.049 \\
    Y1    & 12k   & 0.293 & 0.000 & 0.031 \\
    Y2    & 4e    & 0.000 & 0.000 & 0.096 \\
    \end{tabular}%
  \label{tab:14Hi-structure}%
\end{table}

\begin{table}[htbp]
  \centering
  \caption{DFT relaxed atomic positions for the Mg$_{59}$Y$_8$Zn$_6$ 18R-i LPSO structure, with spacegroup C2/m (12) and lattice parameters a=11.15$\rm\AA$ b=19.34$\rm\AA$ c=16.08$\rm\AA$ $\beta$=76.49$^{\circ}$.}
    \begin{tabular}{llccc}
        \hline\hline
Atom & site & x & y & z \\
\hline
    Mg1   & 8j    & 0.059 & 0.918 & 0.918 \\
    Mg2   & 8j    & 0.053 & 0.752 & 0.917 \\
    Mg3   & 8j    & 0.056 & 0.583 & 0.916 \\
    Mg4   & 8j    & 0.306 & 0.832 & 0.918 \\
    Mg5   & 8j    & 0.305 & 0.665 & 0.919 \\
    Mg6   & 8j    & 0.084 & 0.834 & 0.751 \\
    Mg7   & 8j    & 0.084 & 0.670 & 0.756 \\
    Mg8   & 8j    & 0.330 & 0.915 & 0.756 \\
    Mg9   & 8j    & 0.330 & 0.748 & 0.751 \\
    Mg10  & 8j    & 0.840 & 0.915 & 0.756 \\
    Mg11  & 8j    & 0.191 & 0.828 & 0.586 \\
    Mg12  & 8j    & 0.956 & 0.918 & 0.586 \\
    Mg13  & 8j    & 0.938 & 0.755 & 0.586 \\
    Mg14  & 4i    & 0.310 & 0.000 & 0.918 \\
    Mg15  & 4i    & 0.803 & 0.000 & 0.916 \\
    Mg16  & 4i    & 0.089 & 0.000 & 0.751 \\
 Mg17 int & 2d    & 0.000 & 0.500 & 0.500 \\
    Zn1   & 8j    & 0.427 & 0.888 & 0.614 \\
    Zn2   & 4i    & 0.760 & 0.000 & 0.615 \\
    Y1    & 4j    & 0.170 & 0.647 & 0.573 \\
    Y2    & 4i    & 0.574 & 0.000 & 0.724 \\
    Y3    & 4i    & 0.232 & 0.000 & 0.572 \\
    \end{tabular}%
  \label{tab:18Ri-structure}%
\end{table}

\begin{table}[htbp]
  \centering
  \caption{DFT relaxed lattice parameters for the Mg-X$\rm^L$-Zn LPSO structures, in $\rm\AA$.}
    \begin{tabular}{llccccrcc}
        \hline\hline
                && \multicolumn{4}{c}{18R-i} && \multicolumn{2}{c}{14H-i} \\
\cline{3-6}\cline{8-9}
&X$\rm^L$&    a & b & c & $\beta [^{\circ}]$   && a & c \\
\hline
&	Sc	&	10.99	&	19.05	&	15.84	&	76.52	&	&	11.00	&	35.94	\\
&	Y	&	11.15	&	19.34	&	16.08	&	76.49	&	&	11.15	&	36.36	\\
Exp.\cite{Egusa2012}&   Y   &   11.1    &   19.3    &   16.0    &   76.5   &   &   11.1    &   36.5    \\
&	La	&	11.33	&	19.65	&	16.33	&	76.32	&	&	11.31	&	36.80	\\
&	Ce	&	11.31	&	19.61	&	16.29	&	76.33	&	&	11.30	&	36.73	\\
&	Pr	&	11.28	&	19.56	&	16.25	&	76.35	&	&	11.27	&	36.67	\\
&	Nd	&	11.25	&	19.51	&	16.23	&	76.38	&	&	11.24	&	36.63	\\
&	Pm	&	11.24	&	19.48	&	16.19	&	76.38	&	&	11.23	&	36.56	\\
&	Sm	&	11.21	&	19.44	&	16.18	&	76.41	&	&	11.21	&	36.54	\\
&	Eu	&	11.31	&	19.64	&	16.36	&	76.41	&	&	11.31	&	36.95	\\
&	Gd	&	11.17	&	19.38	&	16.11	&	76.42	&	&	11.18	&	36.45	\\
&	Tb	&	11.16	&	19.36	&	16.09	&	76.42	&	&	11.16	&	36.42	\\
&	Dy	&	11.15	&	19.33	&	16.07	&	76.47	&	&	11.15	&	36.38	\\
&	Ho	&	11.13	&	19.31	&	16.06	&	76.45	&	&	11.15	&	36.39	\\
&	Er	&	11.12	&	19.28	&	16.03	&	76.46	&	&	11.13	&	36.33	\\
&	Tm	&	11.10	&	19.25	&	16.02	&	76.48	&	&	11.11	&	36.29	\\
&	Yb	&	11.24	&	19.49	&	16.26	&	76.48	&	&	11.22	&	36.72	\\
&	Lu	&	11.08	&	19.21	&	15.99	&	76.49	&	&	11.09	&	36.27	\\
	\hline														
&	Tl	&	11.03	&	19.17	&	16.09	&	76.85	&	&	11.04	&	36.56	\\
&	Sb	&	11.06	&	19.13	&	15.96	&	76.73	&	&	11.06	&	36.26	\\
&	Pb	&	11.09	&	19.22	&	16.12	&	76.74	&	&	11.08	&	36.68	\\
&	Na	&	11.10	&	19.23	&	16.16	&	76.62	&	&	11.10	&	36.61	\\
&	Te	&	11.09	&	19.13	&	16.35	&	76.54	&	&	11.06	&	37.12	\\
&	Bi	&	11.15	&	19.29	&	16.10	&	76.55	&	&	11.12	&	36.56	\\
&	Pa	&	11.11	&	19.25	&	16.01	&	76.56	&	&	11.10	&	36.27	\\
&	Ca	&	11.24	&	19.50	&	16.24	&	76.46	&	&	11.23	&	36.72	\\
&	Th	&	11.25	&	19.49	&	16.14	&	76.51	&	&	11.23	&	36.51	\\
&	K	&	11.51	&	19.90	&	16.62	&	76.62	&	&	11.41	&	37.70	\\
&	Sr	&	11.42	&	19.80	&	16.44	&	76.46	&	&	11.40	&	37.11	\\
    \end{tabular}%
  \label{tab:params-zn}%
\end{table}

\begin{table}[htbp]
  \centering
  \caption{DFT relaxed lattice parameters for the Mg-X$\rm^L$-Al LPSO structures, in $\rm\AA$.}
    \begin{tabular}{llccccrcc}
        \hline\hline
                && \multicolumn{4}{c}{18R-i} && \multicolumn{2}{c}{14H-i} \\
\cline{3-6}\cline{8-9}
&X$\rm^L$&    a & b & c & $\beta [^{\circ}]$   && a & c \\
\hline
&	Sc	&	11.03	&	19.11	&	15.90	&	76.58	&	&	11.04	&	36.04	\\
&	Y	&	11.21	&	19.41	&	16.10	&	76.47	&	&	11.19	&	36.42	\\
&	La	&	11.41	&	19.75	&	16.32	&	76.36	&	&	11.37	&	36.80	\\
&	Ce	&	11.39	&	19.71	&	16.29	&	76.36	&	&	11.35	&	36.75	\\
&	Pr	&	11.35	&	19.65	&	16.25	&	76.38	&	&	11.32	&	36.69	\\
&	Nd	&	11.33	&	19.61	&	16.23	&	76.40	&	&	11.30	&	36.61	\\
&	Pm	&	11.30	&	19.57	&	16.20	&	76.43	&	&	11.27	&	36.58	\\
&	Sm	&	11.28	&	19.53	&	16.18	&	76.44	&	&	11.26	&	36.54	\\
&	Eu	&	11.42	&	19.81	&	16.42	&	76.41	&	&	11.39	&	37.02	\\
&	Gd	&	11.24	&	19.46	&	16.14	&	76.48	&	&	11.23	&	36.48	\\
&	Tb	&	11.21	&	19.42	&	16.11	&	76.48	&	&	11.21	&	36.45	\\
&	Dy	&	11.20	&	19.40	&	16.10	&	76.50	&	&	11.20	&	36.44	\\
&	Ho	&	11.19	&	19.37	&	16.09	&	76.50	&	&	11.18	&	36.41	\\
&	Er	&	11.17	&	19.36	&	16.08	&	76.53	&	&	11.17	&	36.39	\\
&	Tm	&	11.16	&	19.34	&	16.07	&	76.55	&	&	11.16	&	36.37	\\
&	Yb	&	11.32	&	19.63	&	16.30	&	76.49	&	&	11.29	&	36.82	\\
&	Lu	&	11.13	&	19.30	&	16.05	&	76.56	&	&	11.13	&	36.35	\\
	\hline														
&	Tl	&	11.03	&	19.13	&	16.19	&	76.94	&	&	11.03	&	36.80	\\
&	Sb	&	11.07	&	19.19	&	16.14	&	76.81	&	&	11.07	&	36.58	\\
&	Pb	&	11.14	&	19.30	&	16.10	&	76.61	&	&	11.13	&	36.52	\\
&	Na	&	11.17	&	19.35	&	16.19	&	76.62	&	&	11.15	&	36.71	\\
&	Te	&	11.10	&	19.26	&	16.44	&	77.26	&	&	11.13	&	37.12	\\
&	Bi	&	11.14	&	19.30	&	16.16	&	76.72	&	&	11.12	&	36.69	\\
&	Pa	&	11.16	&	19.32	&	16.09	&	76.60	&	&	11.15	&	36.45	\\
&	Ca	&	11.38	&	19.71	&	16.37	&	76.49	&	&	11.30	&	36.81	\\
&	Th	&	11.32	&	19.59	&	16.21	&	76.55	&	&	11.29	&	36.65	\\
&	K	&	11.67	&	20.20	&	16.52	&	76.64	&	&	11.55	&	37.48	\\
&	Sr	&	11.50	&	19.96	&	16.50	&	76.41	&	&	11.46	&	37.19	\\
    \end{tabular}%
  \label{tab:params-al}%
\end{table}

\begin{table}[htbp]
  \centering
  \caption{DFT relaxed lattice parameters for the Mg-X$\rm^L$-Cu LPSO structures, in $\rm\AA$.}
    \begin{tabular}{llccccrcc}
        \hline\hline
                && \multicolumn{4}{c}{18R-i} && \multicolumn{2}{c}{14H-i} \\
\cline{3-6}\cline{8-9}
&X$\rm^L$&    a & b & c & $\beta [^{\circ}]$   && a & c \\
\hline
&	Sc	&	10.94	&	18.96	&	15.77	&	76.55	&	&	10.96	&	35.80	\\
&	Y	&	11.08	&	19.22	&	16.03	&	76.55	&	&	11.09	&	36.25	\\
&	La	&	11.23	&	19.49	&	16.23	&	76.35	&	&	11.25	&	36.72	\\
&	Ce	&	11.22	&	19.49	&	16.23	&	76.36	&	&	11.22	&	36.64	\\
&	Pr	&	11.19	&	19.42	&	16.18	&	76.39	&	&	11.20	&	36.58	\\
&	Nd	&	11.17	&	19.39	&	16.16	&	76.43	&	&	11.17	&	36.49	\\
&	Pm	&	11.15	&	19.35	&	16.13	&	76.47	&	&	11.71	&	38.26	\\
&	Sm	&	11.13	&	19.32	&	16.11	&	76.48	&	&	11.15	&	36.43	\\
&	Eu	&	11.22	&	19.46	&	16.28	&	76.53	&	&	11.20	&	36.87	\\
&	Gd	&	11.09	&	19.25	&	16.06	&	76.52	&	&	11.11	&	36.32	\\
&	Tb	&	11.08	&	19.22	&	16.04	&	76.53	&	&	11.10	&	36.30	\\
&	Dy	&	11.08	&	19.21	&	16.03	&	76.56	&	&	11.09	&	36.26	\\
&	Ho	&	11.06	&	19.18	&	16.00	&	76.56	&	&	11.08	&	36.23	\\
&	Er	&	11.05	&	19.15	&	15.98	&	76.57	&	&	11.07	&	36.21	\\
&	Tm	&	11.03	&	19.14	&	15.96	&	76.58	&	&	11.06	&	36.17	\\
&	Yb	&	11.13	&	19.31	&	16.19	&	76.60	&	&	11.12	&	36.69	\\
&	Lu	&	11.02	&	19.10	&	15.93	&	76.55	&	&	11.04	&	36.12	\\
	\hline														
&	Tl	&	10.93	&	18.96	&	15.94	&	76.70	&	&	10.98	&	36.14	\\
&	Sb	&	10.94	&	18.98	&	15.86	&	76.62	&	&	10.96	&	36.01	\\
&	Pb	&	10.97	&	19.01	&	16.05	&	76.94	&	&	10.99	&	36.43	\\
&	Na	&	11.04	&	19.11	&	16.00	&	76.67	&	&	11.03	&	36.41	\\
&	Te	&	11.00	&	19.04	&	16.13	&	76.74	&	&	11.01	&	36.59	\\
&	Bi	&	11.00	&	19.07	&	16.03	&	76.70	&	&	11.02	&	36.38	\\
&	Pa	&	11.03	&	19.10	&	15.91	&	76.51	&	&	11.04	&	36.12	\\
&	Ca	&	11.17	&	19.37	&	16.23	&	76.60	&	&	11.14	&	36.70	\\
&	Th	&	11.16	&	19.34	&	16.08	&	76.47	&	&	11.15	&	36.37	\\
&	K	&	11.39	&	19.72	&	16.60	&	76.71	&	&	11.33	&	37.63	\\
&	Sr	&	11.31	&	19.61	&	16.40	&	76.56	&	&	11.29	&	37.13	\\
    \end{tabular}%
  \label{tab:params-cu}%
\end{table}

\begin{table}[htbp]
  \centering
  \caption{DFT relaxed lattice parameters for the Mg-X$\rm^L$-Co LPSO structures, in $\rm\AA$.}
    \begin{tabular}{llccccrcc}
        \hline\hline
                && \multicolumn{4}{c}{18R-i} && \multicolumn{2}{c}{14H-i} \\
\cline{3-6}\cline{8-9}
&X$\rm^L$&    a & b & c & $\beta [^{\circ}]$   && a & c \\
\hline
&	Sc	&	10.91	&	18.91	&	15.73	&	76.60	&	&	10.94	&	35.78	\\
&	Y	&	11.03	&	19.12	&	15.96	&	76.61	&	&	11.03	&	36.25	\\
&	La	&	11.16	&	19.31	&	16.14	&	76.57	&	&	11.14	&	36.55	\\
&	Ce	&	11.15	&	19.31	&	16.15	&	76.57	&	&	11.14	&	36.58	\\
&	Pr	&	11.12	&	19.26	&	16.10	&	76.58	&	&	11.12	&	36.50	\\
&	Nd	&	11.12	&	19.26	&	16.09	&	76.57	&	&	11.11	&	36.48	\\
&	Pm	&	11.10	&	19.22	&	16.05	&	76.59	&	&	11.09	&	36.42	\\
&	Sm	&	11.06	&	19.17	&	16.01	&	76.58	&	&	11.07	&	36.35	\\
&	Eu	&	11.02	&	19.08	&	16.04	&	76.76	&	&	11.11	&	36.71	\\
&	Gd	&	11.06	&	19.17	&	16.00	&	76.59	&	&	11.05	&	36.27	\\
&	Tb	&	11.03	&	19.11	&	15.95	&	76.58	&	&	11.03	&	36.24	\\
&	Dy	&	11.02	&	19.10	&	15.94	&	76.58	&	&	11.02	&	36.21	\\
&	Ho	&	11.01	&	19.09	&	15.92	&	76.58	&	&	11.02	&	36.19	\\
&	Er	&	11.00	&	19.08	&	15.91	&	76.59	&	&	11.01	&	36.17	\\
&	Tm	&	10.99	&	19.05	&	15.88	&	76.58	&	&	11.00	&	36.13	\\
&	Yb	&	11.06	&	19.15	&	16.07	&	76.70	&	&	11.05	&	36.47	\\
&	Lu	&	10.97	&	19.02	&	15.86	&	76.60	&	&	10.98	&	36.07	\\
	\hline														
&	Tl	&	10.84	&	18.80	&	15.77	&	76.74	&	&	10.87	&	35.93	\\
&	Sb	&	10.80	&	18.75	&	15.88	&	76.96	&	&	10.86	&	36.14	\\
&	Pb	&	10.85	&	18.82	&	15.94	&	77.09	&	&	10.88	&	36.32	\\
&	Na	&	10.96	&	18.99	&	15.85	&	76.68	&	&	10.98	&	36.09	\\
&	Te	&	10.87	&	18.84	&	15.93	&	76.84	&	&	10.93	&	36.13	\\
&	Bi	&	10.87	&	18.86	&	15.99	&	77.02	&	&	10.92	&	36.40	\\
&	Pa	&	11.01	&	19.05	&	15.85	&	76.45	&	&	11.01	&	36.00	\\
&	Ca	&	11.08	&	19.18	&	16.11	&	76.74	&	&	11.07	&	36.52	\\
&	Th	&	11.12	&	19.26	&	16.02	&	76.41	&	&	11.11	&	36.31	\\
&	K	&	11.33	&	19.63	&	16.58	&	76.82	&	&	11.28	&	37.49	\\
&	Sr	&	11.25	&	19.44	&	16.38	&	76.84	&	&	11.20	&	37.03	\\
    \end{tabular}%
  \label{tab:params-co}%
\end{table}

\begin{table}[htbp]
  \centering
  \caption{DFT relaxed lattice parameters for the Mg-X$\rm^L$-Ni LPSO structures, in $\rm\AA$.}
    \begin{tabular}{llccccrcc}
        \hline\hline
                && \multicolumn{4}{c}{18R-i} && \multicolumn{2}{c}{14H-i} \\
\cline{3-6}\cline{8-9}
&X$\rm^L$&    a & b & c & $\beta [^{\circ}]$   && a & c \\
\hline
&	Sc	&	10.94	&	18.94	&	15.73	&	76.63	&	&	10.94	&	35.75	\\
&	Y	&	11.04	&	19.14	&	15.95	&	76.56	&	&	11.06	&	36.22	\\
&	La	&	11.19	&	19.39	&	16.15	&	76.40	&	&	11.18	&	36.58	\\
&	Ce	&	11.18	&	19.38	&	16.14	&	76.40	&	&	11.17	&	36.53	\\
&	Pr	&	11.15	&	19.33	&	16.10	&	76.40	&	&	11.15	&	36.47	\\
&	Nd	&	11.14	&	19.32	&	16.09	&	76.42	&	&	11.13	&	36.44	\\
&	Pm	&	11.11	&	19.26	&	16.05	&	76.44	&	&	11.11	&	36.37	\\
&	Sm	&	11.09	&	19.23	&	16.02	&	76.46	&	&	11.09	&	36.33	\\
&	Eu	&	11.16	&	19.31	&	16.17	&	76.69	&	&	11.69	&	38.55	\\
&	Gd	&	11.07	&	19.19	&	15.99	&	76.50	&	&	11.07	&	36.26	\\
&	Tb	&	11.06	&	19.17	&	15.97	&	76.52	&	&	11.06	&	36.22	\\
&	Dy	&	11.04	&	19.14	&	15.95	&	76.54	&	&	11.05	&	36.19	\\
&	Ho	&	11.03	&	19.12	&	15.93	&	76.55	&	&	11.03	&	36.15	\\
&	Er	&	11.02	&	19.10	&	15.91	&	76.57	&	&	11.03	&	36.15	\\
&	Tm	&	11.01	&	19.09	&	15.90	&	76.59	&	&	11.02	&	36.11	\\
&	Yb	&	11.09	&	19.19	&	16.10	&	76.69	&	&	11.07	&	36.57	\\
&	Lu	&	10.99	&	19.05	&	15.86	&	76.61	&	&	11.01	&	36.08	\\
	\hline														
&	Tl	&	10.85	&	18.80	&	15.87	&	76.78	&	&	10.88	&	36.08	\\
&	Sb	&	10.82	&	18.76	&	15.90	&	76.91	&	&	10.87	&	36.11	\\
&	Pb	&	10.91	&	18.93	&	15.94	&	76.84	&	&	10.94	&	36.31	\\
&	Na	&	11.01	&	19.04	&	15.89	&	76.91	&	&	11.00	&	36.25	\\
&	Te	&	10.88	&	18.85	&	16.00	&	77.40	&	&	10.92	&	36.45	\\
&	Bi	&	10.90	&	18.89	&	16.04	&	76.93	&	&	10.93	&	36.37	\\
&	Pa	&	11.01	&	19.05	&	15.85	&	76.46	&	&	11.01	&	36.03	\\
&	Ca	&	11.09	&	19.20	&	16.09	&	76.69	&	&	11.08	&	36.59	\\
&	Th	&	11.13	&	19.29	&	16.02	&	76.39	&	&	11.12	&	36.26	\\
&	K	&	11.35	&	19.64	&	16.55	&	76.83	&	&	11.31	&	37.50	\\
&	Sr	&	11.27	&	19.48	&	16.35	&	76.70	&	&	11.22	&	37.05	\\
    \end{tabular}%
  \label{tab:params-ni}%
\end{table} 

\begin{figure}[tbp]
\centering %
\includegraphics[width=5in]{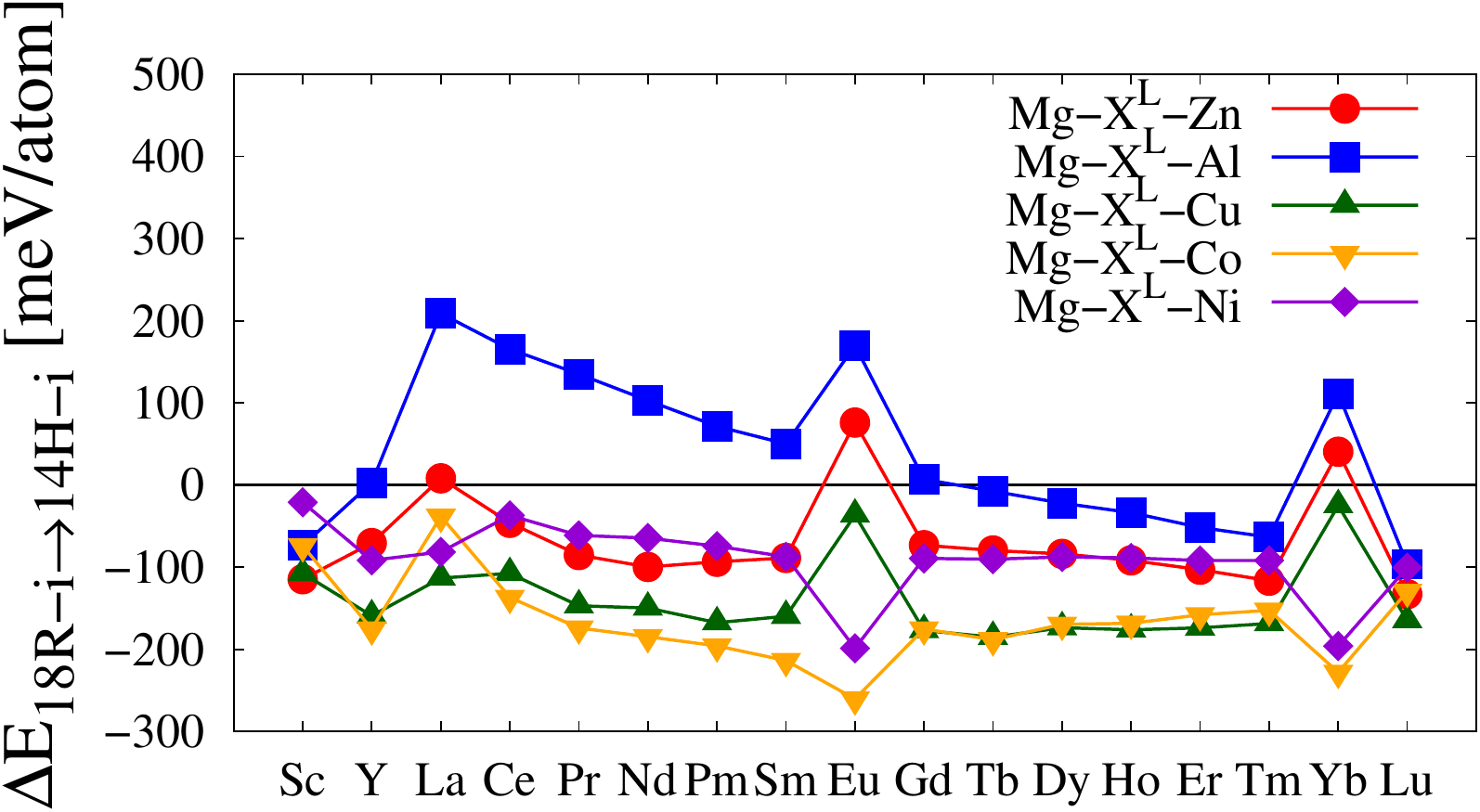}\\
\caption{DFT predicted energy for the transformation between the 18R-i and 14H-i LPSO structures (Equation \ref{eqn:18Ri-14Hi}), $\Delta E\rm_{\operatorname{18R-i}\rightarrow\operatorname{14H-i}}$. Negative values indicate the 14H-i structure is energetically preferred over 18R-i.}
\label{fig:18Ri-14Hi}
\end{figure}

In precipitation experiments, LPSO systems are often observed to initially form the 18R structure and then transform to 14H after annealing\cite{Itoi2004,Kawamura2007}. Mg-Gd-Al is a notable exception, where only the 18R structure has been observed\cite{Yokobayashi2011}. In our previous work, we showed that our calculations are consistent with experiments for the Mg-Y-Zn system, where the 14H structure is more stable than 18R and Mg\cite{Saal2012}. A corresponding relationship between the 14H-i and 18R-i structures is given by the following transformation:
\begin{equation}\label{eqn:18Ri-14Hi}
\rm{2Mg_{59}X^L_8X^S_6[\operatorname{18R-i}]+12Mg\rightarrow Mg_{71}X^L_{8}X^S_{6}[\operatorname{14H-i}]}
\end{equation}
\noindent The DFT predicted energy for this transformation, $\Delta E\rm_{\operatorname{18R-i}\rightarrow\operatorname{14H-i}}$, for every RE-containing LPSO system in this work (X$\rm^L$=RE and X$\rm^S$=Zn,Al,Cu,Co,Ni) is shown in Figure \ref{fig:18Ri-14Hi}. A negative value for $\Delta E\rm_{\operatorname{18R-i}\rightarrow\operatorname{14H-i}}$ indicates the 14H-i structure is more stable than 18R-i and Mg. For most of the systems, the 14H-i structure is more stable, in agreement with experimental observation. Furthermore, for the first half of the Mg-RE-Al series, we predict that the 18R-i structure is preferred, consistent with experimental observation of a preference for 18R LPSO formation in the Mg-Gd-Al system\cite{Yokobayashi2011}. This agreement with experiments, where available, is another indirect indication that the interstitial LPSO structure model is accurate and gives us confidence in DFT thermodynamic predictions for cases where no experimental data exists.

\subsection{Thermodynamic Stability of Mg-RE-X$\rm^S$ LPSO Structures}
\begin{figure}[tbp]
\centering %
\includegraphics[width=5in]{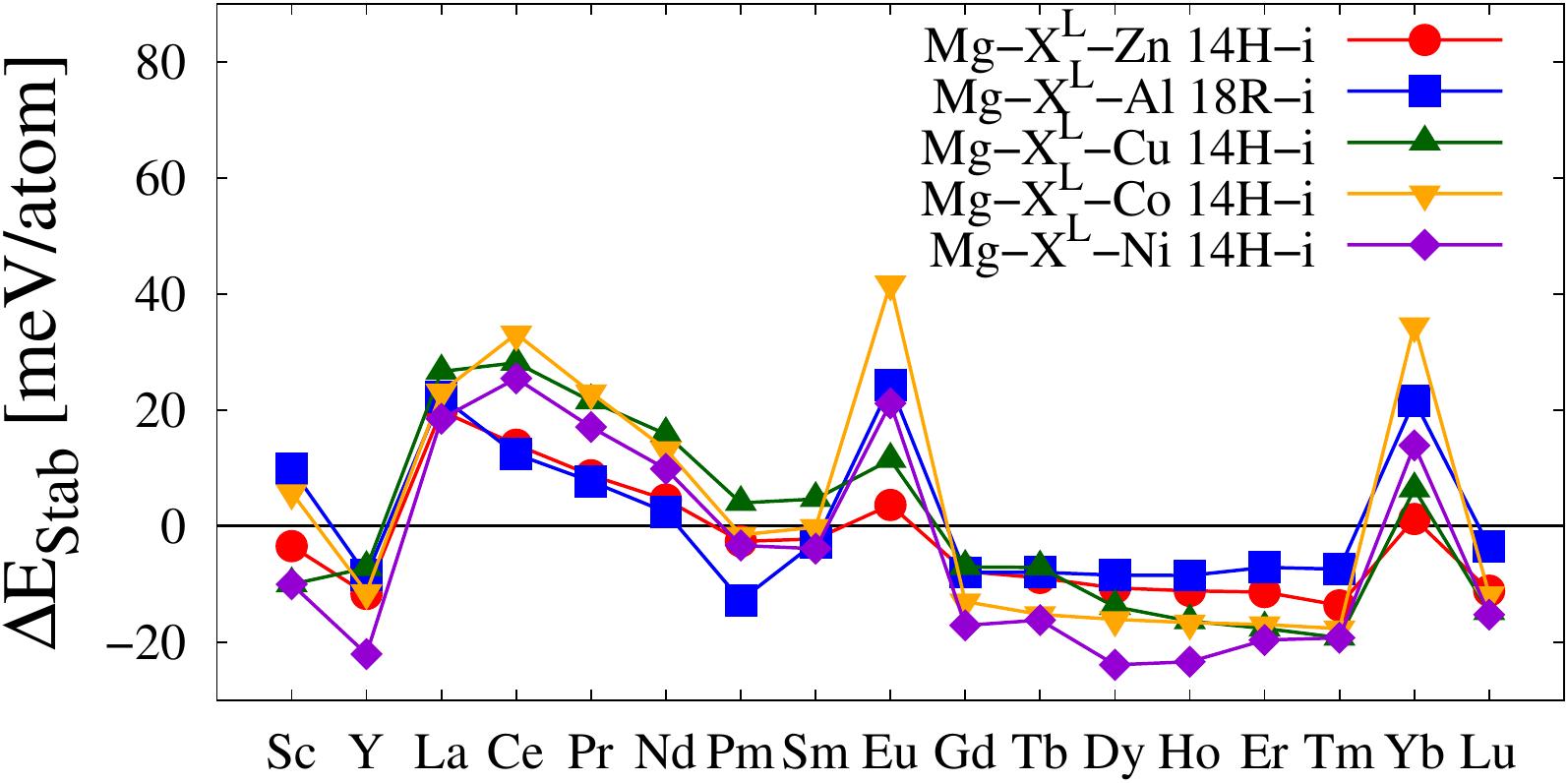}\\
\caption{DFT predicted relative stability of the indicated LPSO structure with respect to the lowest energy combination of all phases known from the ICSD and prototypes database in their respective ternary systems, $\rm{\Delta E_{stab}}$. Negative values indicate the LPSO structure is thermodynamically stable. The sets of stable phases at the LPSO compositions can be found in Tables \ref{tab:Mg-RE-Zn}-\ref{tab:Mg-RE-Ni}.}
\label{fig:stability}
\end{figure}

The formation energies ($\rm{\Delta E_{F}}$) and stabilities ($\rm{\Delta E_{stab}}$) of the Mg-RE-X$\rm^S$ LPSO structures are summarized in Figure \ref{fig:stability}. Nearly all Mg-RE-X$\rm^S$ LPSO phases have negative formation energies, indicating they are stable with respect to the elements --- only the Mg-Eu-Co and Mg-Yb-Co LPSO formation energies are positive. However, a negative formation energy is not a sufficient condition for an LPSO structure to be stable. The LPSO structure must also be more stable than any combination of every other phase in the ternary system, as quantified by $\rm{\Delta E_{stab}}$. To predict $\rm{\Delta E_{stab}}$ of the LPSO structures, we determine the most stable set of competing phases at the 18R-i Mg$_{59}$X$\rm^L_8$X$\rm^S_6$ and 14H-i Mg$_{71}$X$\rm^L_{8}$X$\rm^S_{6}$ compositions. These phases are provided in Tables \ref{tab:Mg-RE-Zn}-\ref{tab:Mg-RE-Ni}. Several 14H-i structures (and 18R-i for X$\rm^S$=Al) have negative values of $\rm{\Delta E_{stab}}$, indicating they are thermodynamically stable, including Mg-Y-Zn. This stability is in contrast to our previous work\cite{Saal2012} where, for 14H Mg-Y-Zn LPSO without the interstitial, the structure lies 11 meV/atom above the convex hull. 14H-i Mg-Y-Zn, in this work, is 12 meV/atom \textit{below} the convex hull. \emph{Thus, using the new interstitial crystal structure\cite{Egusa-PC}, DFT predicts that LPSO structures, in many cases, are thermodynamic ground states.}


\begin{table}[htbp]
  \centering
  \caption{Formation energies and stabilities for the Mg-X$\rm^L$-Zn LPSO structures, in meV/atom. The stable convex hull compounds is given in order of decreasing phase fraction. The number for ICSD compound or the Strukturbericht designation for the simple ordered compounds is given in parentheses. The compounds are the same for both the 18R-i Mg$_{59}$X$\rm^L_8$Zn$_6$ and 14H-i Mg$_{71}$X$\rm^L_{8}$Zn$_{6}$ compositions, unless indicated otherwise by a footnote. A negative stability indicates the LPSO structure is more stable than the convex hull phases.}
    \begin{mpsupertabular}{lccrccl}
        \hline\hline
                & \multicolumn{2}{c}{18R-i} && \multicolumn{2}{c}{14H-i} &\\
\cline{2-3}\cline{5-6}
    X$\rm^L$    & $\Delta$E$\rm_F$ & $\Delta$E$\rm_{stab}$& & $\Delta$E$\rm_F$ & $\Delta$E$\rm_{stab}$ & Convex Hull Phases \\
\hline
Sc	&	-77	&	-4	&	&	-66	&	-3	&	Mg(A3/HCP),ScZn(B2),Mg$_{3}$Sc(D0$_{19}$)	\\
Y	&	-98	&	-13	&	&	-85	&	-12	&	Mg(A3/HCP),MgYZn(160907),Mg$_{3}$Y(D0$_3$)	\\
La	&	-86	&	23	&	&	-74	&	20	&	Mg$_{12}$La(168466),MgLaZn$_{2}$(Heusler),Mg(A3/HCP)\footnote{18R-i: Mg$_{12}$La(168466),MgLaZn$_{2}$(Heusler),Mg$_{3}$La(D0$_3$)}	\\
Ce	&	-88	&	16	&	&	-76	&	14	&	Mg$_{12}$Ce(621495),MgCeZn$_{2}$(Heusler),Mg(A3/HCP)\footnote{18R-i: Mg$_{12}$Ce(621495),MgCeZn$_{2}$(Heusler),Mg$_{41}$Ce$_{5}$(621487)}	\\
Pr	&	-91	&	10	&	&	-78	&	9	&	Mg$_{12}$Pr(104856),MgPrZn$_{2}$(Heusler),Mg(A3/HCP)\footnote{18R-i: Mg$_{12}$Pr(104856),MgPrZn$_{2}$(Heusler),Mg$_{41}$Pr$_{5}$(642771)}	\\
Nd	&	-92	&	6	&	&	-79	&	5	&	Mg$_{41}$Nd$_{5}$(642680),Mg(A3/HCP),MgNdZn$_{2}$(Heusler)	\\
Pm	&	-93	&	-2	&	&	-81	&	-3	&	Mg(A3/HCP),Mg$_{3}$Pm(D0$_{22}$),MgPmZn$_{2}$(Heusler)	\\
Sm	&	-93	&	-2	&	&	-80	&	-2	&	Mg$_{41}$Sm$_{5}$(642842),Mg(A3/HCP),MgSmZn$_{2}$(Heusler)	\\
Eu	&	-79	&	4	&	&	-67	&	4	&	Mg(A3/HCP),Mg$_{2}$Eu(412689),MgEuZn$_{2}$(Heusler)	\\
Gd	&	-92	&	-8	&	&	-80	&	-8	&	Mg(A3/HCP),Mg$_{3}$Gd(D0$_3$),MgGdZn$_{2}$(Heusler)	\\
Tb	&	-91	&	-10	&	&	-79	&	-9	&	Mg(A3/HCP),Mg$_{3}$Tb(D0$_3$),MgTbZn$_{2}$(Heusler)	\\
Dy	&	-90	&	-12	&	&	-78	&	-11	&	Mg(A3/HCP),Mg$_{3}$Dy(D0$_3$),MgDyZn$_{2}$(Heusler)	\\
Ho	&	-88	&	-13	&	&	-76	&	-11	&	Mg(A3/HCP),Mg$_{3}$Ho(D0$_3$),MgHoZn$_{2}$(Heusler)	\\
Er	&	-86	&	-13	&	&	-74	&	-11	&	Mg(A3/HCP),Mg$_{24}$Er$_{5}$(109136),MgErZn$_{2}$(Heusler)	\\
Tm	&	-83	&	-15	&	&	-72	&	-14	&	Mg(A3/HCP),Mg$_{3}$Tm(D0$_3$),MgTmZn$_{2}$(Heusler)	\\
Yb	&	-70	&	1	&	&	-60	&	1	&	Mg(A3/HCP),Mg$_{2}$Yb(104895),YbZn$_{2}$(106234)	\\
Lu	&	-77	&	-12	&	&	-67	&	-11	&	Mg(A3/HCP),LuZn(B2),Mg$_{24}$Lu$_{5}$(642418)	\\
\hline												
Tl	&	-6	&	38	&	&	-5	&	33	&	Mg(A3/HCP),Mg$_{3}$Tl(D0$_{19}$),Mg$_{21}$Zn$_{25}$(240047)\\	
Sb	&	-35	&	86	&	&	-30	&	74	&	Mg(A3/HCP),Mg$_{3}$Sb$_{2}$(2142),Mg$_{21}$Zn$_{25}$(240047)	\\
Pb	&	-13	&	40	&	&	-10	&	36	&	Mg(A3/HCP),Mg$_{3}$Pb(L1$_2$),Mg$_{21}$Zn$_{25}$(240047)\\	
Na	&	17	&	36	&	&	14	&	31	&	Mg(A3/HCP),Mg$_{21}$Zn$_{25}$(240047),Na(C19)	\\
Te	&	-52	&	165	&	&	-45	&	141	&	Mg(A3/HCP),MgTe(52363),Mg$_{21}$Zn$_{25}$(240047)	\\
Bi	&	-27	&	58	&	&	-23	&	50	&	Mg(A3/HCP),Mg$_{3}$Bi$_{2}$(659569),Mg$_{21}$Zn$_{25}$(240047)	\\
Pa	&	66	&	85	&	&	56	&	73	&	Mg(A3/HCP),Mg$_{21}$Zn$_{25}$(240047),Pa(A1/FCC)	\\
Ca	&	-71	&	-3	&	&	-60	&	-2	&	Mg(A3/HCP),CaMg$_{2}$(165564),CaZn$_{2}$(58945)	\\
Th	&	-49	&	-11	&	&	-42	&	-9	&	Mg(A3/HCP),Th$_{2}$Zn(653254),MgThZn$_{2}$(Heusler)	\\
K	&	75	&	94	&	&	67	&	84	&	Mg(A3/HCP),Mg$_{21}$Zn$_{25}$(240047),K(A2/BCC)	\\
Sr	&	-43	&	19	&	&	-37	&	16	&	Mg$_{23}$Sr$_{6}$(104876),Mg(A3/HCP),Mg$_{21}$Zn$_{25}$(240047)	\\
    \end{mpsupertabular}%
  \label{tab:Mg-RE-Zn}%
\end{table}

\begin{table}[htbp]
  \centering
  \caption{Formation energies and stabilities for the Mg-X$\rm^L$-Al LPSO structures, in meV/atom. The stable convex hull compounds is given in order of decreasing phase fraction. The number for ICSD compound or the Strukturbericht designation for the simple ordered compounds is given in parentheses. The compounds are the same for both the 18R-i Mg$_{59}$X$\rm^L_8$Al$_6$ and 14H-i Mg$_{71}$X$\rm^L_{8}$Al$_{6}$ compositions, unless indicated otherwise by a footnote. A negative stability indicates the LPSO structure is more stable than the convex hull phases.}
    \begin{mpsupertabular}{lccrccl}
        \hline\hline
                & \multicolumn{2}{c}{18R-i} && \multicolumn{2}{c}{14H-i} &\\
\cline{2-3}\cline{5-6}
    X$\rm^L$    & $\Delta$E$\rm_F$ & $\Delta$E$\rm_{stab}$& & $\Delta$E$\rm_F$ & $\Delta$E$\rm_{stab}$ & Convex Hull Phases \\
\hline
Sc	&	-76	&	10	&	&	-66	&	7	&	Mg(A3/HCP),AlSc(B2),MgAlSc$_{2}$(Heusler)	\\
Y	&	-101	&	-8	&	&	-87	&	-7	&	Mg(A3/HCP),MgAlY(160908),Mg$_{3}$Y(D0$_3$)	\\
La	&	-93	&	22	&	&	-78	&	21	&	Mg$_{12}$La(168466),Mg(A3/HCP),Al$_{2}$La(57933)\footnote{18R-i: Mg$_{12}$La(168466),Al$_{2}$La(57933),Mg$_{3}$La(D0$_3$)}	 \\
Ce	&	-96	&	12	&	&	-81	&	12	&	Mg$_{12}$Ce(621495),Mg(A3/HCP),Al$_{2}$Ce(57555)\footnote{18R-i: Mg$_{12}$Ce(621495),Al$_{2}$Ce(57555),Mg$_{41}$Ce$_{5}$(621487)}	\\
Pr	&	-98	&	8	&	&	-84	&	7	&	Mg$_{12}$Pr(104856),Mg(A3/HCP),Al$_{2}$Pr(150504)\footnote{18R-i: Mg$_{12}$Pr(104856),Al$_{2}$Pr(150504),Mg$_{41}$Pr$_{5}$(642771)}	\\
Nd	&	-100	&	2	&	&	-85	&	3	&	Mg$_{41}$Nd$_{5}$(642680),Mg(A3/HCP),Al$_{2}$Nd(58027)	\\
Pm	&	-101	&	-13	&	&	-86	&	-10	&	Mg(A3/HCP),Mg$_{3}$Pm(D0$_{22}$),Al$_{3}$Pm(D0$_{19}$)	\\
Sm	&	-100	&	-3	&	&	-85	&	-2	&	Mg$_{41}$Sm$_{5}$(642842),Mg(A3/HCP),Al$_{2}$Sm(58161)	\\
Eu	&	-58	&	24	&	&	-49	&	21	&	Mg(A3/HCP),Mg$_{2}$Eu(412689),Al$_{2}$Eu(57783)	\\
Gd	&	-98	&	-8	&	&	-84	&	-7	&	Mg(A3/HCP),Mg$_{3}$Gd(D0$_3$),Al$_{2}$Gd(57868)	\\
Tb	&	-96	&	-8	&	&	-82	&	-7	&	Mg(A3/HCP),Mg$_{3}$Tb(D0$_3$),Al$_{2}$Tb(58174)	\\
Dy	&	-93	&	-8	&	&	-80	&	-7	&	Mg(A3/HCP),Mg$_{3}$Dy(D0$_3$),Al$_{2}$Dy(107648)	\\
Ho	&	-91	&	-9	&	&	-78	&	-8	&	Mg(A3/HCP),Mg$_{3}$Ho(D0$_3$),Al$_{2}$Ho(57911)	\\
Er	&	-87	&	-7	&	&	-75	&	-7	&	Mg(A3/HCP),Mg$_{24}$Er$_{5}$(109136),Al$_{2}$Er(57764)	\\
Tm	&	-82	&	-7	&	&	-71	&	-7	&	Mg(A3/HCP),Mg$_{3}$Tm(D0$_3$),Al$_{2}$Tm(58192)	\\
Yb	&	-47	&	22	&	&	-40	&	19	&	Mg(A3/HCP),Mg$_{2}$Yb(104895),Al$_{2}$Yb(58223)	\\
Lu	&	-75	&	-4	&	&	-65	&	-4	&	Mg(A3/HCP),Mg$_{24}$Lu$_{5}$(642418),Al$_{2}$Lu(57958)	\\
\hline												
Tl	&	25	&	54	&	&	21	&	46	&	Mg(A3/HCP),Mg$_{3}$Tl(D0$_{19}$),Mg$_{17}$Al$_{12}$(23607)	\\
Sb	&	-5	&	102	&	&	-4	&	88	&	Mg(A3/HCP),Mg$_{3}$Sb$_{2}$(2142),Mg$_{17}$Al$_{12}$(23607)	\\
Pb	&	17	&	56	&	&	15	&	48	&	Mg(A3/HCP),Mg$_{3}$Pb(L1$_2$),Mg$_{17}$Al$_{12}$(23607)	\\
Na	&	45	&	50	&	&	39	&	43	&	Mg(A3/HCP),Mg$_{17}$Al$_{12}$(23607),Na(C19)	\\
Te	&	-17	&	185	&	&	-14	&	160	&	Mg(A3/HCP),MgTe(52363),Mg$_{17}$Al$_{12}$(23607)	\\
Bi	&	3	&	73	&	&	3	&	63	&	Mg(A3/HCP),Mg$_{3}$Bi$_{2}$(659569),Mg$_{17}$Al$_{12}$(23607)	\\
Pa	&	53	&	85	&	&	45	&	72	&	Mg(A3/HCP),AlPa$_{3}$(D0$_{22}$),Al$_{3}$Pa(D0$_{19}$)	\\
Ca	&	-55	&	16	&	&	-47	&	14	&	Mg(A3/HCP),CaMg$_{2}$(165564),CaAl$_{2}$(30213)	\\
Th	&	-55	&	2	&	&	-47	&	2	&	Mg(A3/HCP),AlTh$_{2}$(58180),Al$_{2}$Th(15447)	\\
K	&	104	&	109	&	&	92	&	96	&	Mg(A3/HCP),Mg$_{17}$Al$_{12}$(23607),K(A2/BCC)	\\
Sr	&	-29	&	30	&	&	-23	&	27	&	Mg(A3/HCP),Mg$_{23}$Sr$_{6}$(104876),SrAl$_{2}$(58166)	\\
    \end{mpsupertabular}%
  \label{tab:Mg-RE-Al}%
\end{table}

\begin{table}[htbp]
  \centering
  \caption{Formation energies and stabilities for the Mg-X$\rm^L$-Cu LPSO structures, in meV/atom. The stable convex hull compounds is given in order of decreasing phase fraction. The number for ICSD compound or the Strukturbericht designation for the simple ordered compounds is given in parentheses. The compounds are the same for both the 18R-i Mg$_{59}$X$\rm^L_8$Cu$_6$ and 14H-i Mg$_{71}$X$\rm^L_{8}$Cu$_{6}$ compositions, unless indicated otherwise by a footnote. A negative stability indicates the LPSO structure is more stable than the convex hull phases.}
    \begin{mpsupertabular}{lccrccl}
        \hline\hline
                & \multicolumn{2}{c}{18R-i} && \multicolumn{2}{c}{14H-i} &\\
\cline{2-3}\cline{5-6}
    X$\rm^L$    & $\Delta$E$\rm_F$ & $\Delta$E$\rm_{stab}$& & $\Delta$E$\rm_F$ & $\Delta$E$\rm_{stab}$ & Convex Hull Phases \\
\hline
Sc	&	-67	&	-11	&	&	-58	&	-10	&	Mg(A3/HCP),CuSc(B2),Mg$_{3}$Sc(D0$_{19}$)	\\
Y	&	-84	&	-7	&	&	-73	&	-7	&	Mg(A3/HCP),Mg$_{4}$CuY(419475),Mg$_{3}$Y(D0$_3$)	\\
La	&	-72	&	28	&	&	-62	&	27	&	Mg$_{12}$La(168466),Mg$_{2}$Cu(659334),Mg$_{3}$La(D0$_3$)	\\
Ce	&	-70	&	29	&	&	-61	&	28	&	Mg$_{41}$Ce$_{5}$(621487),Mg$_{2}$Cu(659334),Mg$_{3}$Ce(D0$_3$)	\\
Pr	&	-73	&	22	&	&	-63	&	22	&	Mg$_{41}$Pr$_{5}$(642771),Mg$_{2}$Cu(659334),Mg$_{3}$Pr(104854)	\\
Nd	&	-75	&	16	&	&	-65	&	16	&	Mg$_{41}$Nd$_{5}$(642680),Mg$_{2}$Cu(659334),Mg$_{3}$Nd(D0$_{22}$)	\\
Pm	&	-77	&	6	&	&	-67	&	4	&	Mg(A3/HCP),Mg$_{3}$Pm(D0$_{22}$),Mg$_{2}$Cu(659334)	\\
Sm	&	-77	&	5	&	&	-67	&	5	&	Mg$_{41}$Sm$_{5}$(642842),Mg$_{2}$Cu(659334),Mg$_{3}$Sm(D0$_{22}$)	\\
Eu	&	-67	&	13	&	&	-58	&	11	&	Mg(A3/HCP),Mg$_{2}$Eu(412689),Mg$_{2}$Cu(659334)	\\
Gd	&	-79	&	-7	&	&	-69	&	-7	&	Mg(A3/HCP),Mg$_{3}$Gd(D0$_3$),Mg$_{2}$Cu(659334)	\\
Tb	&	-79	&	-6	&	&	-69	&	-7	&	Mg(A3/HCP),Mg$_{4}$CuTb(418215),Mg$_{3}$Tb(D0$_3$)	\\
Dy	&	-79	&	-15	&	&	-69	&	-14	&	Mg(A3/HCP),Mg$_{3}$Dy(D0$_3$),Mg$_{2}$Cu(659334)	\\
Ho	&	-78	&	-18	&	&	-68	&	-16	&	Mg(A3/HCP),Mg$_{3}$Ho(D0$_3$),Mg$_{2}$Cu(659334)	\\
Er	&	-78	&	-20	&	&	-68	&	-18	&	Mg$_{24}$Er$_{5}$(109136),Mg(A3/HCP),Mg$_{2}$Cu(659334)	\\
Tm	&	-76	&	-21	&	&	-66	&	-19	&	Mg(A3/HCP),CuTm(B2),Mg$_{3}$Tm(D0$_3$)	\\
Yb	&	-61	&	8	&	&	-53	&	6	&	Mg(A3/HCP),Mg$_{2}$Yb(104895),Mg$_{2}$Cu(659334)	\\
Lu	&	-73	&	-16	&	&	-64	&	-15	&	Mg(A3/HCP),CuLu(B2),Mg$_{24}$Lu$_{5}$(642418)	\\
\hline												
Tl	&	-3	&	53	&	&	-2	&	46	&	Mg(A3/HCP),Mg$_{3}$Tl(D0$_{19}$),Mg$_{2}$Cu(659334)	\\
Sb	&	-34	&	99	&	&	-27	&	87	&	Mg(A3/HCP),Mg$_{3}$Sb$_{2}$(2142),Mg$_{2}$Cu(659334)	\\
Pb	&	-12	&	53	&	&	-11	&	45	&	Mg(A3/HCP),Mg$_{3}$Pb(L1$_2$),Mg$_{2}$Cu(659334)	\\
Na	&	34	&	65	&	&	29	&	56	&	Mg(A3/HCP),Mg$_{2}$Cu(659334),Na(C19)	\\
Te	&	-50	&	179	&	&	-42	&	154	&	Mg(A3/HCP),Mg$_{2}$Cu(659334),MgTe(52363)	\\
Bi	&	-24	&	73	&	&	-19	&	64	&	Mg(A3/HCP),Mg$_{3}$Bi$_{2}$(659569),Mg$_{2}$Cu(659334)	\\
Pa	&	67	&	98	&	&	56	&	83	&	Mg(A3/HCP),Mg$_{2}$Cu(659334),Pa(A1/FCC)	\\
Ca	&	-57	&	19	&	&	-49	&	16	&	Mg(A3/HCP),CaMg$_{2}$(165564),Mg$_{2}$Cu(659334)	\\
Th	&	-35	&	-4	&	&	-31	&	-4	&	Mg(A3/HCP),Mg$_{2}$Cu(659334),Th(A1/FCC)	\\
K	&	89	&	120	&	&	79	&	106	&	Mg(A3/HCP),Mg$_{2}$Cu(659334),K(A2/BCC)	\\
Sr	&	-28	&	45	&	&	-22	&	41	&	Mg$_{23}$Sr$_{6}$(104876),Mg(A3/HCP),Mg$_{2}$Cu(659334)	\\
    \end{mpsupertabular}%
  \label{tab:Mg-RE-Cu}%
\end{table}

\begin{table}[htbp]
  \centering
  \caption{Formation energies and stabilities for the Mg-X$\rm^L$-Co LPSO structures, in meV/atom. The stable convex hull compounds is given in order of decreasing phase fraction. The number for ICSD compound or the Strukturbericht designation for the simple ordered compounds is given in parentheses. The compounds are the same for both the 18R-i Mg$_{59}$X$\rm^L_8$Co$_6$ and 14H-i Mg$_{71}$X$\rm^L_{8}$Co$_{6}$ compositions, unless indicated otherwise by a footnote. A negative stability indicates the LPSO structure is more stable than the convex hull phases.}
    \begin{mpsupertabular}{lccrccl}
        \hline\hline
                & \multicolumn{2}{c}{18R-i} && \multicolumn{2}{c}{14H-i} &\\
\cline{2-3}\cline{5-6}
    X$\rm^L$    & $\Delta$E$\rm_F$ & $\Delta$E$\rm_{stab}$& & $\Delta$E$\rm_F$ & $\Delta$E$\rm_{stab}$ & Convex Hull Phases \\
\hline
Sc	&	-63	&	6	&	&	-54	&	6	&	Mg(A3/HCP),CoSc(B2),Mg$_{3}$Sc(D0$_{19}$)	\\
Y	&	-61	&	-12	&	&	-53	&	-11	&	Mg(A3/HCP),Mg$_{3}$Y(D0$_3$),Co$_{3}$Y(625559)	\\
La	&	-50	&	23	&	&	-43	&	23	&	Mg$_{12}$La(168466),Mg$_{3}$La(D0$_3$),Co$_{13}$La(656879)	\\
Ce	&	-37	&	36	&	&	-33	&	33	&	Mg$_{41}$Ce$_{5}$(621487),Mg$_{12}$Ce(621495),Co(A3/HCP)\footnote{18R-i: Mg$_{41}$Ce$_{5}$(621487),Co(A3/HCP),Mg$_{3}$Ce(D0$_3$)}	\\
Pr	&	-43	&	25	&	&	-38	&	23	&	Mg$_{41}$Pr$_{5}$(642771),Mg$_{12}$Pr(104856),Co(A3/HCP)\footnote{18R-i: Mg$_{41}$Pr$_{5}$(642771),Co(A3/HCP),Mg$_{3}$Pr(104854)}	\\
Nd	&	-47	&	16	&	&	-42	&	13	&	Mg$_{41}$Nd$_{5}$(642680),Co(A3/HCP),Mg(A3/HCP)\footnote{18R-i: Mg$_{41}$Nd$_{5}$(642680),Co(A3/HCP),Mg$_{3}$Nd(D0$_{22}$)}	 \\
Pm	&	-52	&	0	&	&	-46	&	-2	&	Mg(A3/HCP),Mg$_{3}$Pm(D0$_{22}$),Co(A3/HCP)	\\
Sm	&	-54	&	1	&	&	-47	&	0	&	Mg$_{41}$Sm$_{5}$(642842),Mg(A3/HCP),Co$_{17}$Sm$_{2}$(625233)\footnote{18R-i: Mg$_{41}$Sm$_{5}$(642842),Co$_{17}$Sm$_{2}$(625233),Mg$_{3}$Sm(D0$_{22}$)}	\\
Eu	&	1	&	50	&	&	0	&	42	&	Mg(A3/HCP),Mg$_{2}$Eu(412689),Co(A3/HCP)	\\
Gd	&	-59	&	-13	&	&	-52	&	-13	&	Mg(A3/HCP),Mg$_{3}$Gd(D0$_3$),Co$_{17}$Gd$_{2}$(623333)	\\
Tb	&	-61	&	-17	&	&	-53	&	-15	&	Mg(A3/HCP),Mg$_{3}$Tb(D0$_3$),Co$_{2}$Tb(152587)	\\
Dy	&	-62	&	-18	&	&	-54	&	-16	&	Mg(A3/HCP),Mg$_{3}$Dy(D0$_3$),Co$_{2}$Dy(163700)	\\
Ho	&	-62	&	-18	&	&	-55	&	-17	&	Mg(A3/HCP),Mg$_{3}$Ho(D0$_3$),Co$_{2}$Ho(108296)	\\
Er	&	-63	&	-18	&	&	-55	&	-17	&	Mg(A3/HCP),Mg$_{24}$Er$_{5}$(109136),Co$_{2}$Er(622773)	\\
Tm	&	-63	&	-20	&	&	-55	&	-18	&	Mg(A3/HCP),Mg$_{3}$Tm(D0$_3$),Co$_{2}$Tm(625505)	\\
Yb	&	3	&	41	&	&	2	&	35	&	Mg(A3/HCP),Mg$_{2}$Yb(104895),Co(A3/HCP)	\\
Lu	&	-62	&	-13	&	&	-54	&	-12	&	Mg(A3/HCP),CoLu(B2),CoLu$_{3}$(624053)	\\
\hline												
Tl	&	48	&	72	&	&	40	&	61	&	Mg(A3/HCP),Mg$_{3}$Tl(D0$_{19}$),Co(A3/HCP)	\\
Sb	&	-24	&	78	&	&	-21	&	67	&	Mg(A3/HCP),Mg$_{3}$Sb$_{2}$(2142),Co(A3/HCP)	\\
Pb	&	28	&	62	&	&	23	&	52	&	Mg(A3/HCP),Mg$_{3}$Pb(L1$_2$),Co(A3/HCP)	\\
Na	&	128	&	128	&	&	109	&	109	&	Mg(A3/HCP),Na(C19),Co(A3/HCP)	\\
Te	&	-18	&	180	&	&	-15	&	155	&	Mg(A3/HCP),MgTe(52363),Co(A3/HCP)	\\
Bi	&	2	&	67	&	&	2	&	58	&	Mg(A3/HCP),Mg$_{3}$Bi$_{2}$(659569),Co(A3/HCP)	\\
Pa	&	-25	&	12	&	&	-18	&	13	&	Mg(A3/HCP),Co$_{3}$Pa(L1$_2$),Pa(A1/FCC)	\\
Ca	&	14	&	59	&	&	11	&	49	&	Mg(A3/HCP),CaMg$_{2}$(165564),Co(A3/HCP)	\\
Th	&	-69	&	-6	&	&	-60	&	-6	&	Mg(A3/HCP),CoTh(625442),Co$_{3}$Th$_{7}$(625455)	\\
K	&	184	&	184	&	&	159	&	159	&	Mg(A3/HCP),K(A2/BCC),Co(A3/HCP)	\\
Sr	&	49	&	91	&	&	41	&	77	&	Mg(A3/HCP),Mg$_{23}$Sr$_{6}$(104876),Co(A3/HCP)	\\
    \end{mpsupertabular}%
  \label{tab:Mg-RE-Co}%
\end{table}

\begin{table}[htbp]
  \centering
  \caption{Formation energies and stabilities for the Mg-X$\rm^L$-Ni LPSO structures, in meV/atom. The stable convex hull compounds is given in order of decreasing phase fraction. The number for ICSD compound or the Strukturbericht designation for the simple ordered compounds is given in parentheses. The compounds are the same for both the 18R-i Mg$_{59}$X$\rm^L_8$Ni$_6$ and 14H-i Mg$_{71}$X$\rm^L_{8}$Ni$_{6}$ compositions. A negative stability indicates the LPSO structure is more stable than the convex hull phases.}
    \begin{mpsupertabular}{lccrccl}
        \hline\hline
                & \multicolumn{2}{c}{18R-i} && \multicolumn{2}{c}{14H-i} &\\
\cline{2-3}\cline{5-6}
    X$\rm^L$    & $\Delta$E$\rm_F$ & $\Delta$E$\rm_{stab}$& & $\Delta$E$\rm_F$ & $\Delta$E$\rm_{stab}$ & Convex Hull Phases \\
\hline
Sc	&	-106	&	-12	&	&	-91	&	-10	&	Mg(A3/HCP),NiSc(B2),Mg$_{3}$Sc(D0$_{19}$)	\\
Y	&	-112	&	-25	&	&	-97	&	-22	&	Mg(A3/HCP),Mg$_{3}$Y(D0$_3$),Mg$_{2}$Ni(30713)	\\
La	&	-98	&	18	&	&	-85	&	18	&	Mg$_{12}$La(168466),Mg$_{2}$Ni(30713),Mg$_{3}$La(D0$_3$)	\\
Ce	&	-90	&	25	&	&	-78	&	25	&	Mg$_{41}$Ce$_{5}$(621487),Mg$_{2}$Ni(30713),Mg$_{3}$Ce(D0$_3$)	\\
Pr	&	-95	&	17	&	&	-82	&	17	&	Mg$_{41}$Pr$_{5}$(642771),Mg$_{2}$Ni(30713),Mg$_{3}$Pr(104854)	\\
Nd	&	-99	&	8	&	&	-85	&	10	&	Mg$_{41}$Nd$_{5}$(642680),Mg$_{2}$Ni(30713),Mg$_{3}$Nd(D0$_{22}$)	\\
Pm	&	-102	&	-3	&	&	-88	&	-3	&	Mg(A3/HCP),Mg$_{3}$Pm(D0$_{22}$),Mg$_{2}$Ni(30713)	\\
Sm	&	-104	&	-6	&	&	-90	&	-4	&	Mg$_{41}$Sm$_{5}$(642842),Mg$_{2}$Ni(30713),Mg$_{3}$Sm(D0$_{22}$)	\\
Eu	&	-71	&	25	&	&	-62	&	21	&	Mg(A3/HCP),Mg$_{2}$Eu(412689),Mg$_{2}$Ni(30713)	\\
Gd	&	-109	&	-19	&	&	-94	&	-17	&	Mg(A3/HCP),Mg$_{3}$Gd$_{2}$Ni$_{2}$(421933),Mg$_{3}$Gd(D0$_3$)	\\
Tb	&	-110	&	-18	&	&	-95	&	-16	&	Mg(A3/HCP),Mg$_{3}$Ni$_{2}$Tb$_{2}$(240761),Mg$_{3}$Tb(D0$_3$)	\\
Dy	&	-111	&	-27	&	&	-96	&	-24	&	Mg(A3/HCP),DyNi(109242),Mg$_{3}$Dy(D0$_3$)	\\
Ho	&	-112	&	-27	&	&	-96	&	-23	&	Mg(A3/HCP),HoNi(106792),Mg$_{3}$Ho(D0$_3$)	\\
Er	&	-112	&	-23	&	&	-97	&	-20	&	Mg(A3/HCP),ErNi(630842),Mg$_{24}$Er$_{5}$(109136)	\\
Tm	&	-111	&	-22	&	&	-96	&	-19	&	Mg(A3/HCP),NiTm(105428),Mg$_{3}$Tm(D0$_3$)	\\
Yb	&	-67	&	18	&	&	-59	&	14	&	Mg(A3/HCP),Mg$_{2}$Yb(104895),Mg$_{2}$Ni(30713)	\\
Lu	&	-110	&	-16	&	&	-95	&	-15	&	Mg(A3/HCP),LuNi(642448),Mg$_{24}$Lu$_{5}$(642418)	\\
\hline												
Tl	&	-13	&	59	&	&	-11	&	51	&	Mg(A3/HCP),Mg$_{3}$Tl(D0$_{19}$),Mg$_{2}$Ni(30713)	\\
Sb	&	-60	&	89	&	&	-51	&	77	&	Mg(A3/HCP),Mg$_{3}$Sb$_{2}$(2142),Mg$_{2}$Ni(30713)	\\
Pb	&	-30	&	51	&	&	-26	&	44	&	Mg(A3/HCP),Mg$_{3}$Pb(L1$_2$),Mg$_{2}$Ni(30713)	\\
Na	&	46	&	93	&	&	38	&	79	&	Mg(A3/HCP),Mg$_{2}$Ni(30713),Na(C19)	\\
Te	&	-67	&	178	&	&	-56	&	154	&	Mg(A3/HCP),Mg$_{2}$Ni(30713),MgTe(52363)	\\
Bi	&	-45	&	68	&	&	-39	&	58	&	Mg(A3/HCP),Mg$_{3}$Bi$_{2}$(659569),Mg$_{2}$Ni(30713)	\\
Pa	&	9	&	56	&	&	-10	&	31	&	Mg(A3/HCP),Mg$_{2}$Ni(30713),Pa(A1/FCC)	\\
Ca	&	-58	&	34	&	&	-52	&	27	&	Mg(A3/HCP),CaMg$_{2}$(165564),Mg$_{2}$Ni(30713)	\\
Th	&	-89	&	-13	&	&	-77	&	-12	&	Mg(A3/HCP),NiTh(105403),Ni$_{3}$Th$_{7}$(105406)	\\
K	&	99	&	146	&	&	85	&	126	&	Mg(A3/HCP),Mg$_{2}$Ni(30713),K(A2/BCC)	\\
Sr	&	-26	&	64	&	&	-23	&	54	&	Mg$_{23}$Sr$_{6}$(104876),Mg(A3/HCP),Mg$_{2}$Ni(30713)	\\
    \end{mpsupertabular}%
  \label{tab:Mg-RE-Ni}%
\end{table}

\begin{figure}[tbp]
\centering %
\includegraphics[width=5in]{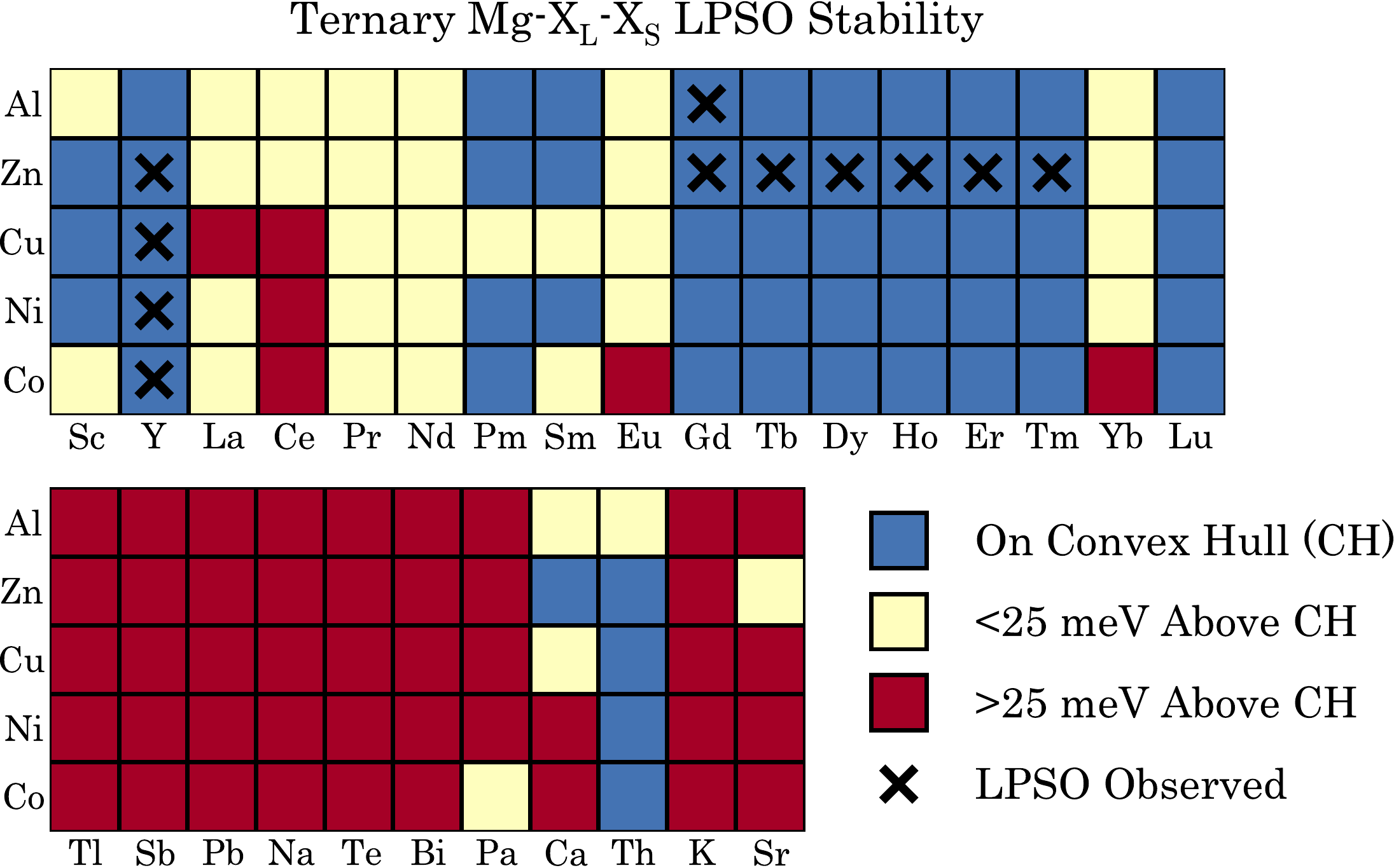}\\
\caption{DFT predicted stability of 14H-i and 18R-i LPSO structures for all Mg-X$\rm^L$-X$\rm^S$ ternary systems explored in this work. X$\rm^S$ and X$\rm^L$ elements are given along the vertical and horizontal axes, respectively. Color coding is defined by the values of $\rm{\Delta E_{stab}}$ given in Tables \ref{tab:Mg-RE-Zn}-\ref{tab:Mg-RE-Ni}: blue for on the convex hull ($\rm{\Delta E_{stab}}<0$), yellow for near the convex hull ($0<\rm{\Delta E_{stab}}<25$ meV/atom), and red for far from the convex hull (25 meV/atom $<\rm{\Delta E_{stab}}$). X$\rm^L$=RE systems are given at top and X$\rm^L\not=$RE systems at bottom. Experimentally observed LPSO-forming systems are also indicated\cite{Amiya2003,Yamasaki2005,Kawamura2006,YAMADA2006,Kawamura2007,Itoi2008,Nie2008,Yokobayashi2011,Mi2013,Jin2013,Leng2013}. Note that blue squares without an ``x'' indicate systems where we predict as-yet-unobserved LPSO phases to be stable.}
\label{fig:stabmap}
\end{figure}

The stability of LPSO structures in all Mg-RE-X$\rm^S$ ternary systems explored in the current work is summarized in Figure \ref{fig:stabmap}. Interestingly, regardless of which X$\rm^S$ is present, the same set of heavier RE X$\rm^L$ elements generally appear to form stable LPSO structures: Y, Gd, Tb, Dy, Ho, Er, Tm, and Lu. As indicated in Figure \ref{fig:stabmap}, several other ternary systems, such as those containing Nd and Sm, are predicted to have \emph{nearly} stable LPSO structures, lying less than 25 meV above the convex hull (k$\rm_B$T at room temperature). Currently, LPSO phases have only been studied in very few ternaries for X$\rm^S\not=$Zn\cite{Kawamura2006,Yokobayashi2011,Itoi2008,Mi2013,Jin2013,Leng2013}.
Of the 85 Mg-X$\rm^S$-RE systems we explored with DFT, 52 are predicted to have thermodynamically stable LPSO structures. To our knowledge, only 11 LPSO-forming ternary systems have been reported in the literature\cite{Amiya2003,Yamasaki2005,Kawamura2006,YAMADA2006,Kawamura2007,Itoi2008,Nie2008,Yokobayashi2011,Mi2013,Jin2013,Leng2013}, and we predict all 11 to contain stable LPSO structures. Therefore, we predict the existence of 41 new, as-yet-unobserved LPSO-forming ternary systems. We call for experimental investigation of these systems to confirm (or deny) the validity of our predictions.


\subsection{Thermodynamic Stability of Non-RE LPSO Structures}
Up to now, we have only discussed LPSO structures in ternary systems where X$\rm^L$=RE. Non-RE X$\rm^L$ elements are highly desirable to reduce the cost of employing LPSO precipitate strengthening on an industrial scale. To predict with DFT every possible Mg-X$\rm^L$-X$\rm^S$ system is prohibitively expensive given the large quantity of possible ternary systems. Therefore, we limit the current DFT exploration of non-RE LPSO systems to the five known X$\rm^S$ elements and employ a simple screen (detailed below) on all possible X$\rm^L$ elements with high-throughput DFT calculations that are less computationally more efficient than full calculations of LPSO stability. The set of promising X$\rm^L$ elements which pass this screen is sufficiently small for DFT predictions of stability to be performed. We conclude by testing the predicted DFT LPSO stabilities with several observed X$\rm^L$ trends for LPSO formation\cite{Kawamura2007} to investigate why only specific X$\rm^L$ elements form stable LPSO structures.

\begin{figure}[tbp]
\centering %
\includegraphics[width=5in]{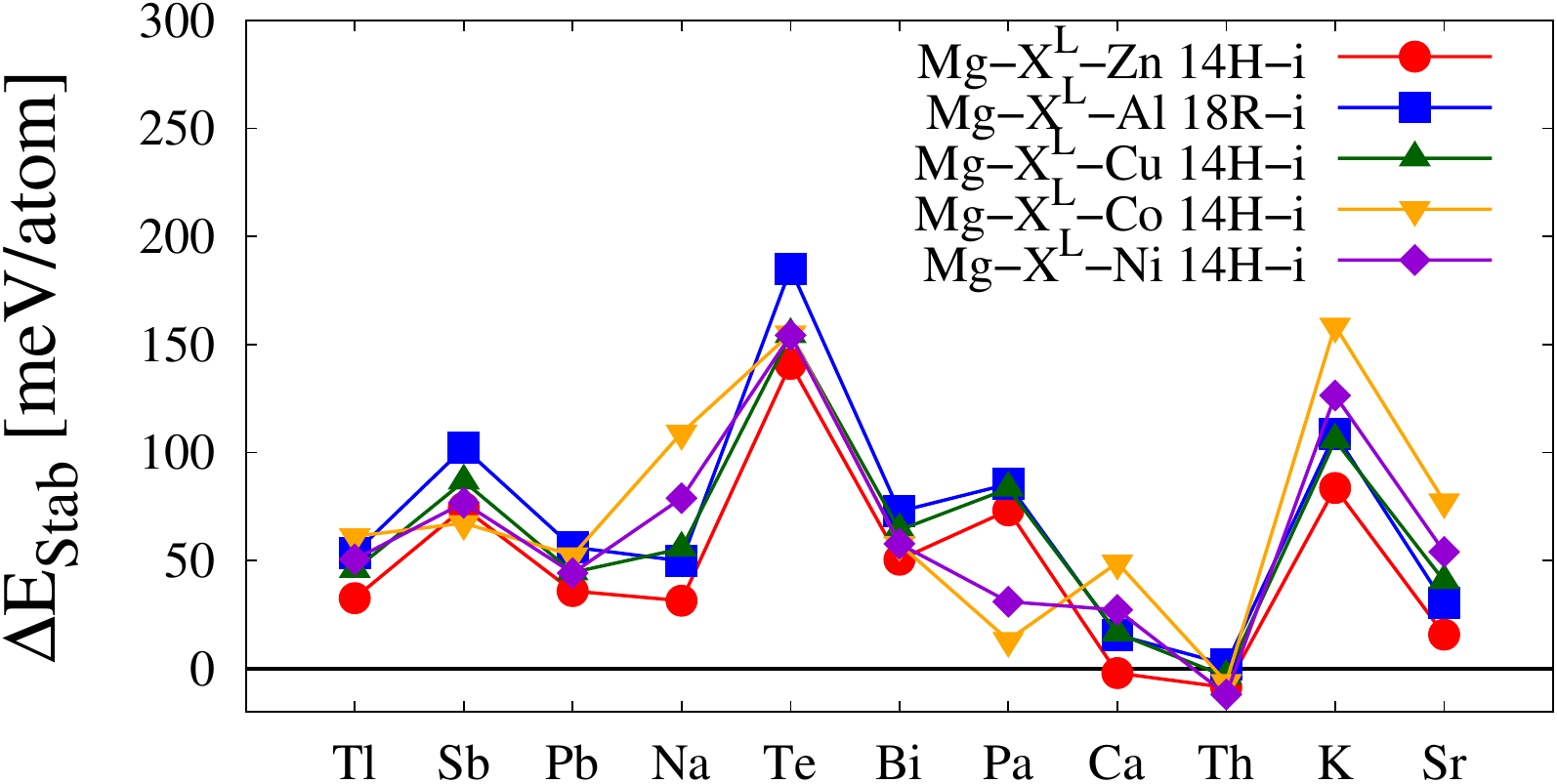}\\
\caption{DFT predicted relative stability of the indicated LPSO structure with respect to the lowest energy combination of all phases known from the ICSD and prototypes database in their respective ternary systems, $\rm{\Delta E_{stab}}$. Negative values indicate the LPSO structure is thermodynamically stable. The sets of stable phases at the LPSO compositions can be found in Tables \ref{tab:Mg-RE-Zn}-\ref{tab:Mg-RE-Ni}. Elements are ordered in increasing impurity volume in Mg.}
\label{fig:stab_other}
\end{figure}

We screen candidate X$\rm^L$ elements for LPSO formation with what we assume to be the most important factor contributing to the ability of an X$\rm^L$ element to form a stable LPSO structure: the size mismatch of the element relative to Mg, noted as an important criteria for X$\rm^L$ LPSO structure formation\cite{Kawamura2007}. To quantify this mismatch, Kawamura et al.\cite{Kawamura2007} used the mismatch between elemental atomic radii. From our DFT predicted atomic radii (calculated by taking half the nearest neighbor distance in the 0K ground state crystal structure), the atomic radius mismatch of the observed X$\rm^L$ elements (Y and the later REs, as given in Figure \ref{fig:stabmap}) range between 8.5-12\% larger than Mg. After calculating this quantity for 88 elements,\footnote{These 88 elements are those with a GGA-PBE potential provided with version 5.3.2 of VASP.} only three have radius mismatches near this range: Pb, Tl, and Th. The stability of LPSO structures for these elements serving as X$\rm^L$ was predicted with DFT. Shown in Figures \ref{fig:stab_other} and \ref{fig:stabmap} and given in Tables \ref{tab:Mg-RE-Zn}-\ref{tab:Mg-RE-Ni}, the stabilities for the Pb- and Tl-containing LPSO structures are very positive, indicating they will not form LPSO structures. Th-containing LPSO structures, on the other hand, are predicted to be stable. However, thorium's radioactive nature presumably limits the usefulness of any potential LPSO structures containing the element.

We find a better indicator of the Mg/X$\rm^L$ size mismatch towards LPSO stability with the impurity volume, $\rm{V_{Imp}^{X^L}}$. This quantity is defined by:
\begin{equation} \label{eqn:vimp}
\rm{V_{Imp}^{X}=V(Mg_{149}X_1)-V(Mg_{150})}
\end{equation}
where $\rm{V(Mg_{150})}$ and $\rm{V(Mg_{149}X_1)}$ are the volumes of a 150 atom HCP supercell containing Mg$_{150}$ and Mg$_{149}$X, respectively. In our previous work, supercells of 150 atoms were found to produce values for $\rm{V_{Imp}}$ converged with respect to supercell size\cite{Saal2011b}. Predicted for many elements in our previous works\cite{Shin2010a,Saal2011b}, the impurity volume of X$\rm^L$ in Mg captures the interaction of the alloying element with the Mg matrix. Currently, we have calculated the DFT impurity volume for every element with a VASP potential. As shown in Figure \ref{fig:design}, $\rm{V_{Imp}^{X^L}}$ , as an LPSO-forming criteria, clusters all the known X$\rm^L$ elements (Y and the later REs, as given in Figure \ref{fig:stabmap}) into a single group (between 11.1 and 14.6 $\rm\AA^3$). Therefore, we predict with DFT the LPSO stability of several non-RE solutes with impurity volumes near RE values, specifically K, Sr, Ca, Na, Sb, Pb, Bi, and Pa. These stabilities are shown in Figure \ref{fig:stab_other} and given in Tables \ref{tab:Mg-RE-Zn}-\ref{tab:Mg-RE-Ni}. Most of these LPSO structures are found to be metastable, but some come energetically close to the T=0K ground state convex hull, as shown in Figure \ref{fig:stabmap}, particularly Ca- and Sr-containing systems. In these systems, finite-temperature effects could stabilize LPSO structures. \emph{Our results suggest that Ca- and Sr-containing systems are promising and should be explored experimentally for possible new non-RE LPSO phases as they lie near the convex hull (2 meV/atom below the hull in the Mg-Ca-Zn system).}

\subsection{Testing Proposed Design Rules for LPSO Stability}
Kawamura et al.\cite{Kawamura2007} observed several trends amongst LPSO-forming X$\rm^L$ elements: (1) X$\rm^L$ is larger than Mg, (2) the mixing enthalpy between Mg/X$\rm^L$ and X$\rm^L$/X$\rm^S$ is favorable, (3) X$\rm^L$ has the HCP structure at room temperature, and (4) X$\rm^L$ is moderately soluble in Mg. The first trend was used as the screening criteria for choosing non-RE elements. With our DFT calculated energetics database of LPSO structures in 85 RE- and 50 non-RE-containing ternary systems, the remaining trends can be examined more closely and used to elucidate why RE X$\rm^L$ elements form stable LPSO structure whereas others do not.

\begin{figure}[tbp]
\centering %
\includegraphics[width=5in]{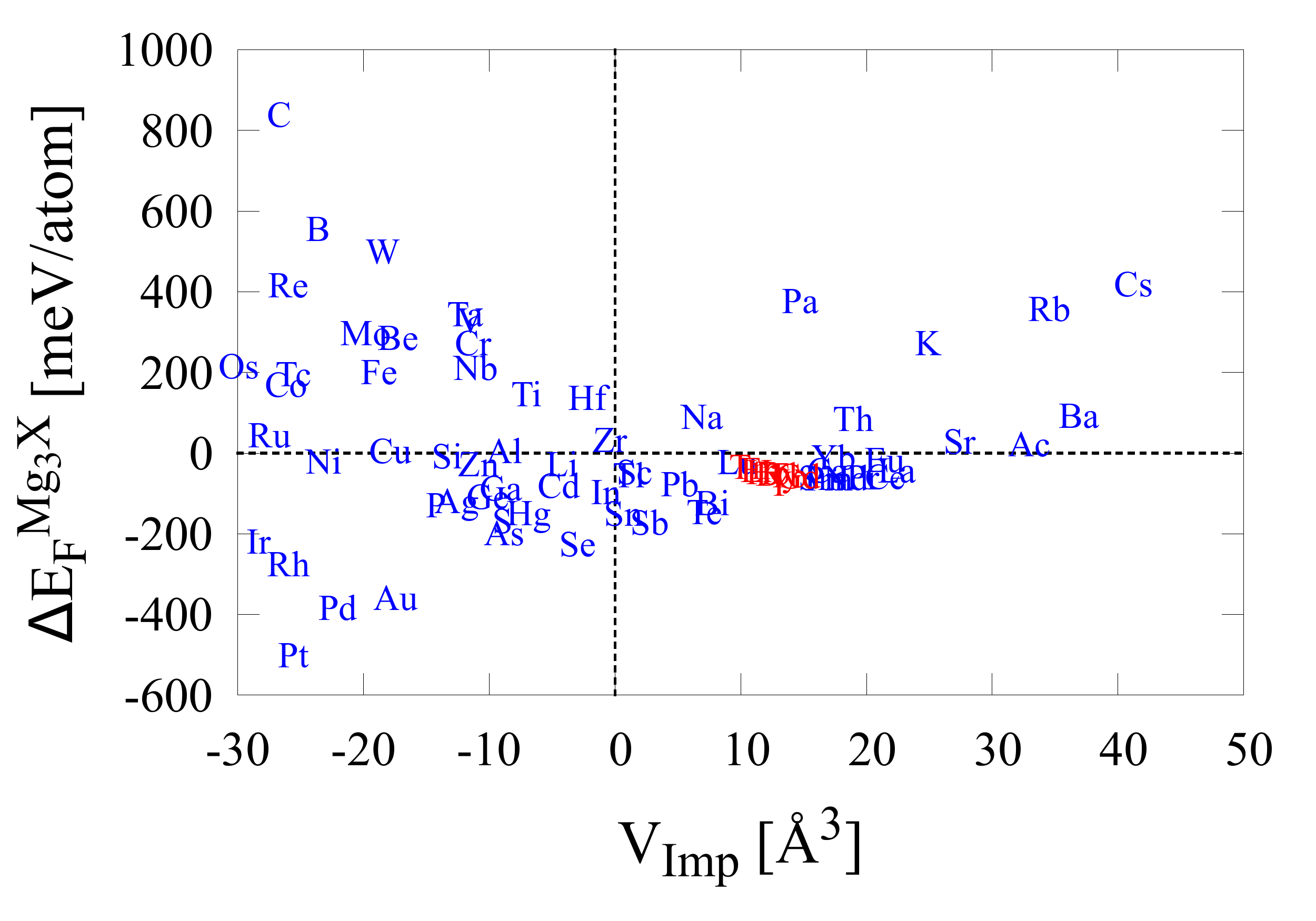}\\
\includegraphics[width=5in]{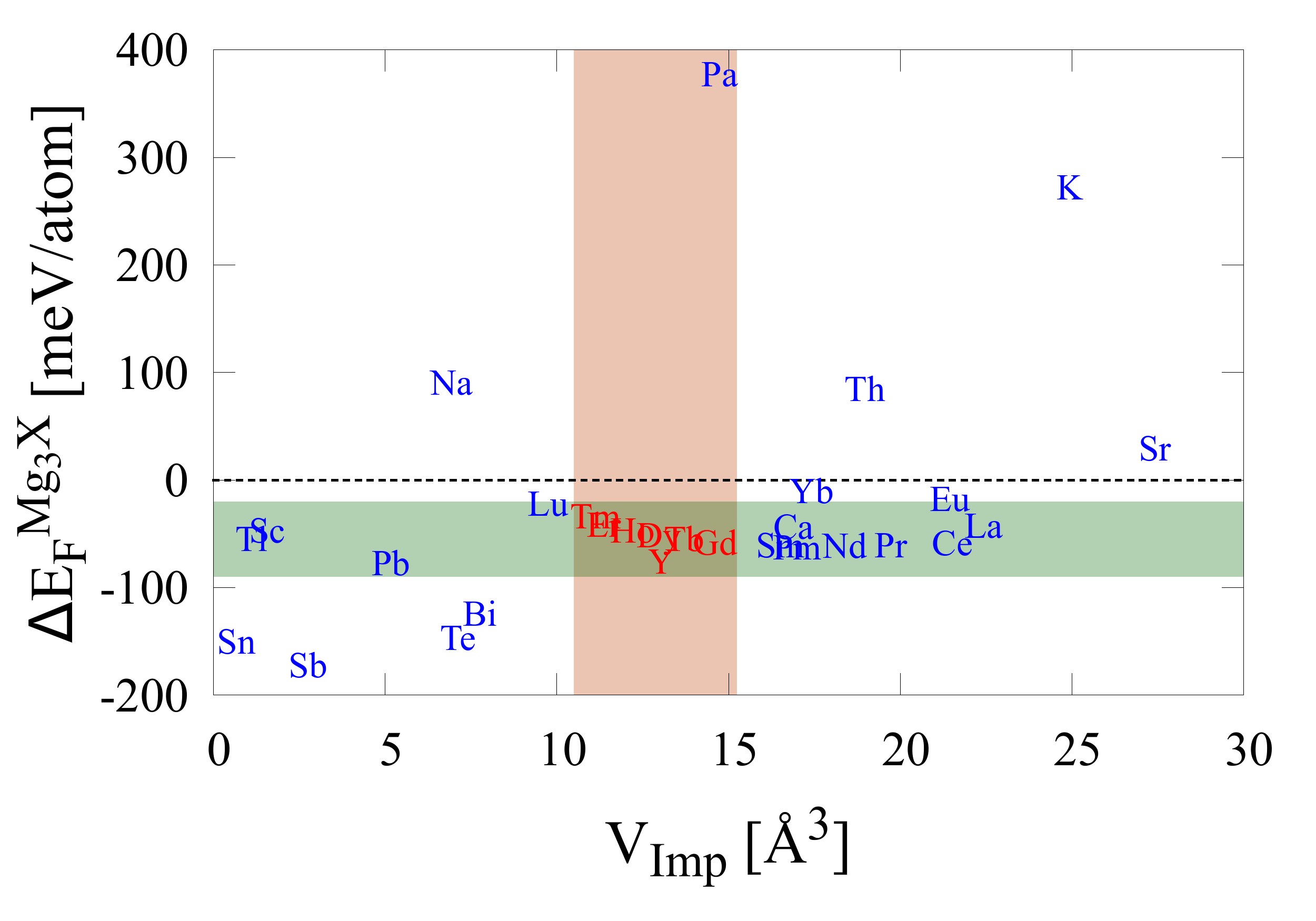}\\
\caption{Formation energy of Mg$_3$X L1$_2$ vs. the impurity volume of X in Mg (top) for all elements tested in this work and (bottom) close up of positive impurity volume region. Observed LPSO-forming X$\rm^L$ elements are in red font. Shaded regions correspond to range of formation energies and impurity volumes for observed LPSO-forming X$\rm^L$ elements. }
\label{fig:design}
\end{figure}

The second proposed trend is that the Mg-X$\rm^L$ and X$\rm^L$-X$\rm^S$ binary systems exhibit favorable mixing thermodynamics. This trend had been observed from results of the Miedema model\cite{Kawamura2007}. The favorable interactions between these elements may promote the formation of the LPSO, as Mg-X$\rm^L$ and X$\rm^L$-X$\rm^S$ nearest-neighbor bonds are present in the binary and ternary layers of the LPSO structure. Unlike predictions of the Miedema model, DFT calculations of the formation energies of simple ordered compounds can estimate binary interactions \emph{for a particular lattice}. As the X$\rm^L$ atoms bond with Mg and X$\rm^S$ on both HCP and FCC lattices in the LPSO structure, L1$_2$ and D0$_{19}$ formation energies for many possible Mg-X$\rm^L$ and X$\rm^L$-X$\rm^S$ systems were calculated with DFT.  The Mg$_3$X L1$_2$ formation energy, $\rm{\Delta E_F^{Mg_3X}}$, appeared to be the best indicator for whether an X$\rm^L$ element can contribute to a stable LPSO structure, by clustering observed X$\rm^L$ elements (Y and the later REs, as given in Figure \ref{fig:stabmap}) with similar values. As seen in Figure \ref{fig:design}, all observed X$\rm^L$ elements have negative Mg$_3$X L1$_2$ formation energies, between -34 and -76 meV/atom.

Interestingly, either $\rm{\Delta E_F^{Mg_3X}}$ or $\rm{V_{Imp}^{X^L}}$ alone are not sufficient indicators of whether an X$\rm^L$ element will form a stable LPSO structure.  For instance, Pb is predicted to have formation energies in the range of the observed X$\rm^L$ elements, but, from Figure \ref{fig:stab_other}, Pb forms metastable LPSO structures.  From Figure \ref{fig:design}, Pb has a smaller impurity volume than the observed RE X$\rm^L$ elements. Pa, conversely, has an impurity volume similar to the observed X$\rm^L$ elements but has a very unfavorable mixing energy, also resulting in metastable LPSO structures. As seen in Figure \ref{fig:design}, of all the non-RE elements studied in this work, Ca is nearest to satisfying both constraints, perhaps explaining why Ca-containing LPSO structures are predicted to have competitive stabilities. Therefore, we find that the impurity volume and X$\rm^L$-Mg FCC mixing energy together serve as excellent criteria for determining LPSO formation, including, within a certain range, all stable X$\rm^L$ elements and excluding all others, as indicated by the intersection of the two shaded regions in Figure \ref{fig:design}. The heavy RE elements are unique in that they satisfy both criteria.

The remaining two trends of Kawamura et al.\cite{Kawamura2007} can be explored from direct experimental observations. The third trend is that all known X$\rm^L$ elements appear to be HCP at room temperature. Every HCP RE element has been found to form LPSO structures, except for Sc and Lu which, to our knowledge, have not been explored. From our DFT results, we predict that Sc- and Lu-containing LPSO structures are stable. Non-RE HCP elements include Be, Ti, Zr, Tc, Ru, Hf, Re, Os, and Tl. From our predictions of the impurity volume, these elements are all smaller than Mg, except for Tl, which is only slightly larger than Mg. With an impurity volume about 90\% smaller than the values for the observed X$\rm^L$ elements (see Figure \ref{fig:design}), Tl is predicted to form metastable LPSO structures (see Figure \ref{fig:stab_other}). This result shows that there are no non-RE HCP elements that also have impurity volumes in the range of the RE elements. Ca, Sr, and Th, which are the promising LPSO forming X$\rm^L$ elements discussed earlier, are not HCP.  However, DFT calculations of HCP Ca and Sr predict it to be very close energetically to FCC Ca and Sr (within 5 meV/atom or less)\cite{Wang2004a,Saal2013}.  The fourth trend is that some moderate degree of solubility of X$\rm^L$ in Mg is present. From the observed X$\rm^L$ elements, the solubility at the eutectic temperature varies between 3.4 and 6.9 at.\%\cite{Villars}. The solubility of Ag lies in this range, but the impurity volume of Ag is negative. Again, Ca and Th do not satisfy these conditions, exhibiting solubilities of 0.44 and 0.52 at.\%, respectively\cite{Villars}.

Ultimately, of the 11 non-RE X$\rm^L$ elements studied in this work, only Ca, Sr, and Th are found to form low-energy stable and/or metastable structures competitive with the thermodynamic ground state, at least with the five X$\rm^S$ elements presently considered. Although we predict Th-containing LPSO structures to have stabilities comparable to those containing REs, Th exhibits an impurity volume and mixing energy that makes it an unlikely LPSO-former. As no Mg-Th-X$\rm^S$ ternary compounds or phase diagrams have been reported\cite{Villars,Bergerhoff1983,Belsky2002}, the prediction of stable Th-containing LPSO structures is likely an indication our DFT database of competing structures is incomplete for Mg-X$\rm^S$-Th ternary systems. Previous studies on the Mg-Ca-Zn system\cite{Somekawa2007,Zhang2010,Oh-ishi2009,Zhou2013} have not found an LPSO structure, although the nominal alloy compositions are different from those for typical LPSO systems (e.g. Mg$_{97}$RE$_2$Zn$_1$\cite{Kawamura2007}). One study\cite{Oh-ishi2009} examined the age hardening response for 0.3 at.\% Ca with varying amounts of Zn and found 0.6 at.\% Zn gives the best response, with peak hardness coinciding with monolayer G.P. zones aligned along basal planes. A recent study\cite{Zhou2013} of Mg-2.4Zn-3.2Ca (at.\%) using rapid solidification/powder metallurgy, a common approach to forming LPSO structures, did not report LPSO formation, although again, the nominal alloy composition is different from other LPSO alloys. With regard to the Mg-Zn-Sr system, no ternary crystal structures or phase diagrams have been reported\cite{Villars,Bergerhoff1983,Belsky2002}. Only recently have Mg alloys containing Zn and Sr been investigated\cite{Guan2013,Brar2012}, as a potential medical implant material. The study by Brar et al.\cite{Brar2012} examined a series of cast alloys at Mg-(0.8-2.3)Zn-0.1Sr (at.\%) compositions, finding Mg-rich ternary phases of composition Mg$_{70}$Zn$_{25}$Sr$_5$ and Mg$_{71}$Zn$_{23}$Sr$_6$. Their crystal structures were not reported. However, a later study by Guan et al.\cite{Guan2013} investigated a Mg-1.5Zn-0.3Sr (at.\%) cast alloy and found only binary precipitates. Even if LPSO structures do not form in these non-RE systems, Ca and Sr may still prove useful as an alloying element to reduce the RE content in LPSO precipitates without dramatically degrading the stability of the phase (e.g. adding Ca to the Mg-Y-Zn LPSO structure may allow reducing the Y-content of the alloy without greatly reducing the volume fraction of LPSO precipitates).

\section{Summary}
We have predicted, with DFT, the properties of many Mg-X$\rm^L$-X$\rm^S$ LPSO structures: 85 systems consisting of X$\rm^L$=RE and X$\rm^S$=Zn,Al,Cu,Co,Ni. We find that the interstitial 14H-i and 18R-i LPSO structure models are energetically preferred to the previously published gradual structure model for every system we tested. Using these interstitial LPSO structures, we demonstrate that LPSO phases are thermodynamically stable in several ternary systems, specifically when X$\rm^L$ is Y or a heavy RE element. We predict LPSO structures to be stable for every system where they are observed. Which X$\rm^L$ elements form stable LPSO structures appear to be generally independent of the five X$\rm^S$ elements in the current work, suggesting that there are several novel LPSO systems to explore, as summarized in Figure \ref{fig:stabmap}.

On the basis of the size mismatch between Mg and X$\rm^L$, quantified by the impurity volume, several potential X$\rm^L$ elements(Tl, Sb, Pb, Na, Te, Bi, Pa, Ca, Th, K, Sr) have been explored, amounting to 50 non-RE ternary LPSO systems. These results have been used to examine previously proposed trends of X$\rm^L$ elements that form LPSO structures, particularly the size mismatch and mixing energy between Mg and X$\rm^L$. We find that the impurity volume and X$\rm^L$-Mg FCC mixing energy together serve as excellent criteria determining LPSO formation, including, within a certain range, all stable X$\rm^L$ elements and excluding all others. Of all explored non-RE LPSO ternary systems, Ca-, Sr-, and Th-containing LPSO structures are the most competitive with the ground state. Ca- and Sr-containing LPSO systems are suggested for further experimental study as an alternative to or addition to RE alloying.

\section*{Acknowledgements}
Funding support provided by the Ford-Boeing-Northwestern Alliance. Calculations were performed on the Northwestern University high performance computing system, Quest. Many thanks to Don Shih, Bita Ghaffari, and Haruyuki Inui for fruitful discussions. We would also like to thank Daisuke Egusa and Eiji Abe at the University of Tokyo for discussions regarding the interstitial LPSO structure.

\newpage
   \begin{singlespace}
   \bibliographystyle{model1a-num-names}
   \bibliography{lpso2}
   \end{singlespace}

\end{document}